# UNIQORN: Unified Question Answering over RDF Knowledge Graphs and Natural Language Text

Soumajit Pramanik[a,1], Jesujoba Alabi[b], Rishiraj Saha Roy[b] and Gerhard Weikum[b]

[a] Indian Institute of Technology Bhilai, Bhilai, India
[b] Max Planck Institute for Informatics, Saarbruecken, Germany

ABSTRACT

Question answering over RDF data like knowledge graphs has been greatly advanced, with a number of good systems providing crisp answers for natural language questions or telegraphic queries. Some of these systems incorporate textual sources as additional evidence for the answering process, but cannot compute answers that are present in text alone. Conversely, the IR and NLP communities have addressed QA over text, but such systems barely utilize semantic data and knowledge. This paper presents a method for *complex questions* that can seamlessly operate over a mixture of RDF datasets and text corpora, or individual sources, in a unified framework. Our method, called UNIQORN, builds a context graph on-the-fly, by retrieving question-relevant evidences from the RDF data and/or a text corpus, using fine-tuned BERT models. The resulting graph typically contains all question-relevant evidences but also a lot of noise. UNIQORN copes with this input by a graph algorithm for Group Steiner Trees, that identifies the best answer candidates in the context graph. Experimental results on several benchmarks of complex questions with multiple entities and relations, show that UNIQORN significantly outperforms state-of-the-art methods for *heterogeneous QA* – in a full training mode, as well as in zero-shot settings. The graph-based methodology provides user-interpretable evidence for the complete answering process.

## 1. Introduction

**Motivation**. Question answering (QA) aims to compute direct answers to information needs posed as natural language (NL) utterances [99, 77, 28, 88, 10, 123, 56]. We focus on the class of *factoid questions* that are objective in nature and have one or more named entities as answers [2, 41, 122, 96, 22, 94]. Early research [98, 123] used patterns to extract text passages with candidate answers, or had sophisticated pipelines like the proprietary IBM WATSON system [45, 46] that won the Jeopardy! quiz show. With the rise of large knowledge graphs (KGs) or knowledge bases (KBs) like YAGO [110], DBpedia [7], Freebase [15], and Wikidata [124], the focus shifted from text corpora to these structured RDF[1] data sources, represented as subject-predicate-object (SPO) triples. We refer to such answering of questions over knowledge graphs as *KG-QA*.

While KGs capture a large part of the world's encyclopedic knowledge, they are inherently incomplete. This is because emerging and ephemeral facts (e.g., teams losing in semi-finals of ongoing sports leagues, or celebrities dating each other) are included only with notable delays or not at all (if deemed unworthy for the KG). Also, user interests may go beyond the predicates that are modeled in KGs like Wikidata. As a result, considering text from the open Web, like news websites and Wikipedia, as an input source, is a necessity for many QA applications. We call this paradigm as question answering over text, or *Text-QA* for short.

This work aims to unify the two paradigms of KG-QA and Text-QA, so that a single method can tap into either KG or text or both simultaneously, depending on the nature of questions. Our design rationale is to treat the KG as the source with the highest quality but limited coverage, and tap text sources as a complement: either to fill in gaps in the KG content or to corroborate candidate answers by multiple sources of evidence. In all these settings, a key requirement observed by our approach is to also provide supporting references for a system-generated answer: KG facts and/or specific text sources.

A timely alternative for this kind of universal QA could perhaps be Large Language Models (LLM) [16, 106, 118]. However, it is in the generative nature of these models that they come with a risk of hallucinations and an inherent lack of providing tangible provenance for answers. Therefore, we do not consider LLM-for-QA as a viable alternative. To shed more light into this issue, we include small-scale experiments with two state-of-the-art LLMs in Sec. 6.3.

**Limitations of the state-of-the-art**. The stark contrast in the representation of content in the sources of KG-QA (structured SPO triples) and Text-QA (natural language sequences) spawned two completely different threads of research in factual question answering. The prevalent paradigms in KG-QA focused on building explicit queries or logical forms that could be executed over the RDF triple store [117, 37, 14, 1, 91, 58, 107], or using approximate graph search techniques after mapping question phrases to KG items [122, 22, 105, 62]. Methods in Text-QA, on the other hand, converged into a retriever-reader model where a set of question relevant passages or evidences are retrieved first, followed by extracting or generating an answer string from these top evidences [6, 141, 26, 19, 78]. The fallout of these parallel branches of research was that methods for one source were completely incompatible with those for the other: it is not very meaningful to apply logical forms or run graph search over text sequences, and building a reader model to extract answers from structured data does not make sense. The need for simultaneously tapping into both sources was recognized and advanced by a suite of algorithms for heterogeneous QA, but most of these methods had ad hoc pipelines for each source, that interacted at various stages to produce a final list of answers [102, 129, 131, 112, 111, 103].

*Corresponding author
pramanik@mpi-inf.mpg.de (S. Pramanik); jalabi@mpi-inf.mpg.de (J. Alabi); rishiraj@mpi-inf.mpg.de (R.S. Roy); weikum@mpi-inf.mpg.de (G. Weikum)
ORCID(s): 0000-0002-2297-1153 (S. Pramanik); 0000-0002-2297-1153 (J. Alabi); 0000-0002-5774-5658 (R.S. Roy); 0000-0003-4959-6098 (G. Weikum)

[1] Most of this work was done while the author was at the Max Planck Institute for Informatics.

[1] https://www.w3.org/RDF/

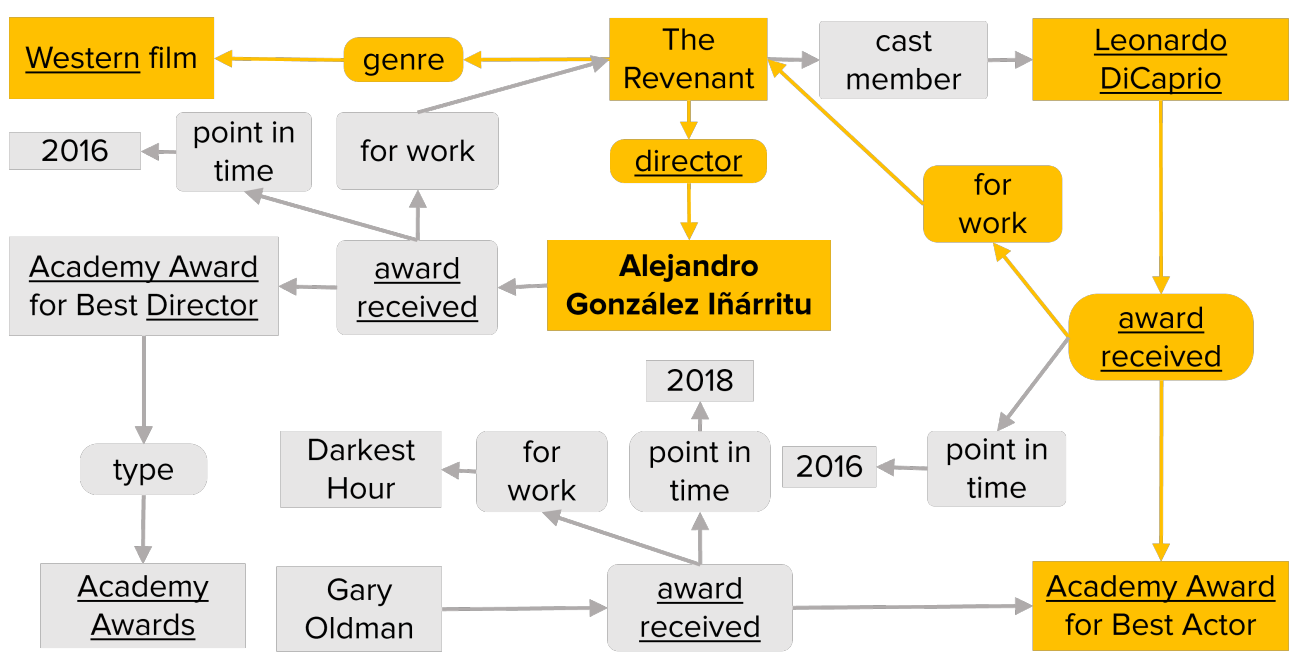

(a) $XG(q)$ example for KG as input.

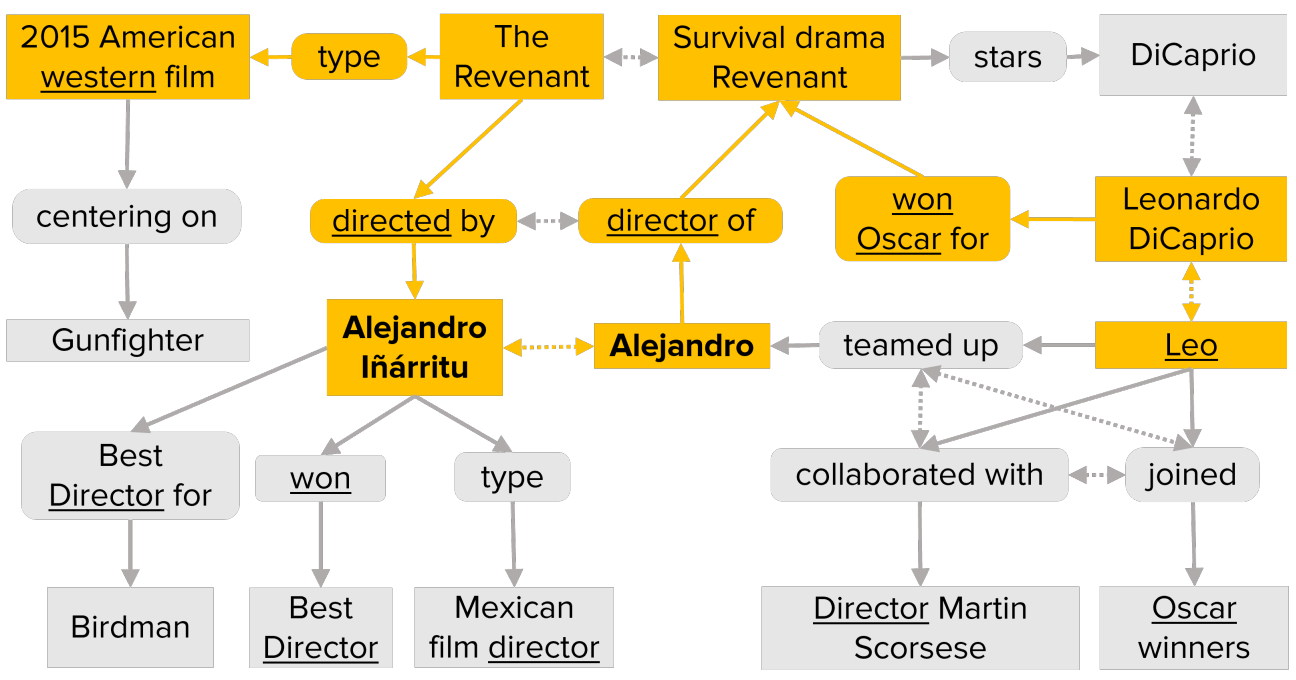

(b) $XG(q)$ example for text as input.

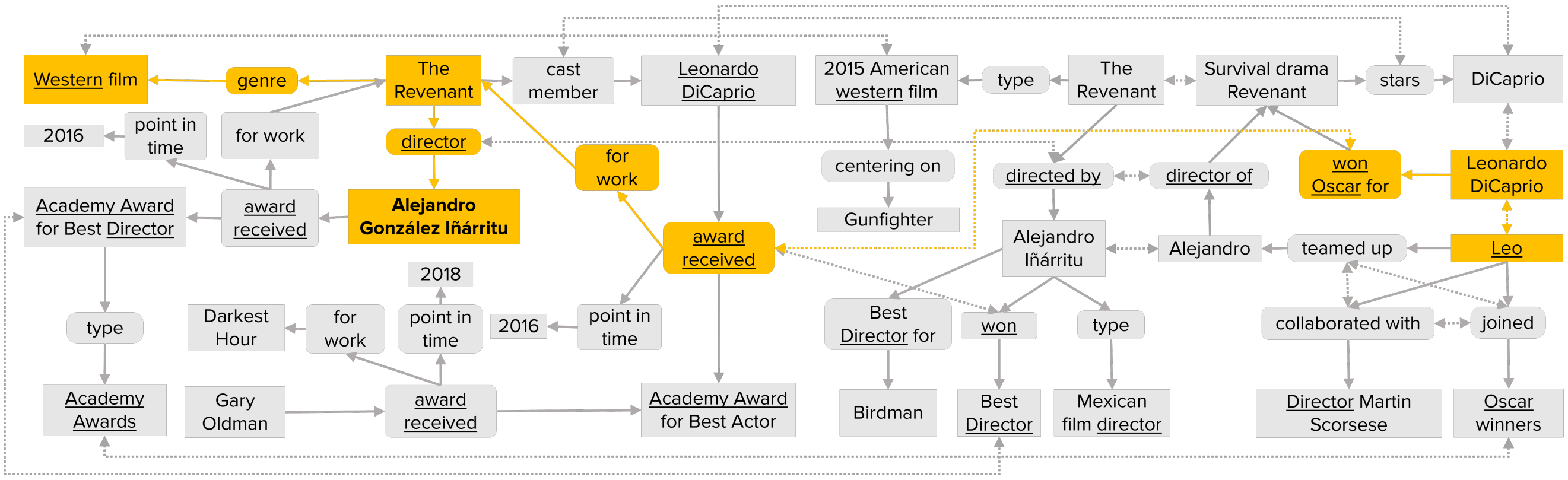

(c) $XG(q)$ example for KG and text as input.

**Figure 1:** Context graphs (XG) built by UNIQORN in each of the answering setups for the question $q$ = *director of the western for which Leo won an Oscar?* Anchors are nodes with parts of their labels underlined (that match with $q$); answers are in bold. Orange subgraphs are Group Steiner Trees. (color online)

Very recently, the UNIK-QA model [89] showed an elegant solution: a unified representation of these structured and unstructured sources via *verbalization* (sometimes referred to as serialization or linearization) of each evidence in the structured data into a text sequence. Once all pieces of evidence are in a unified textual form, state-of-the-art generative models from Text-QA could be applied on a suitably small question-relevant evidence pool. While such an approach works well for answering simple questions, we argue that it has the following shortcomings:

- it cannot make the most of the information contained in text sources for multi-step inference;
- the flattening of the inherent relationships between KG facts through verbalization makes it inadequate for more complex information needs;
- a generative reader model in the final answering step [64] cannot explain the steps for answer derivation and cannot provide users with evidence that supports the answer.

**Approach**. To overcome these limitations, we propose UNIQORN, a Unified framework for question answering over RDF knowledge graphs and Natural language text, that is a method for *complex QA over heterogeneous sources*. Our proposal hinges on two key ideas:

- Handling KG-QA, Text-QA, and heterogeneous setups with the same unified method, by building a noisy KG-like context graph from KG or text inputs on-the-fly for each question, using named entity disambiguation (NED), open information extraction (open IE), and BERT fine-tuning. Conceptually, while UNIK-QA *flattens evidence* from structured sources to create a text sequence, UNIQORN adopts the *opposite philosophy* and *induces structure* from text sequences to unify the sources in a graph setup: this helps us cope with more complex questions.
- We use graph algorithms for Group Steiner Trees [39, 113], where question-relevant cues are connected to produce an answer along with an explanation. Since answers are extracted from pieces of KG or text evidence, they are traceable to their source.

**Table 1**
Concepts and notation.

| Notation | Concept | Notation | Concept |
|---|---|---|---|
| $K$ | Knowledge graph (KG) | $\mathcal{S}$ | Snippet |
| $D$ | Document corpus | $q = \langle q_i \rangle$ | Question with question words |
| $K^D$ | Quasi KG from $D$ | $A = \{a\}$ | Answer(s) to $q$ |
| $K^H$ | Heterogeneous $KG$ | $XG^{(\cdot)}(q)$ | Context graph for $q$ built from $K/K^D/K^H$ |
| $E = \{e\}$ | Entity | $N = \{n\}$ | Node |
| $P = \{p\}$ | Predicate | $\mathcal{E} = \{\epsilon\}$ | Edge |
| $\mathbb{T} = \{t\}$ | Type | $N^{\mathcal{T}}, \mathcal{E}^{\mathcal{T}}$ | Node types, edge types |
| $L = \{l\}$ | Literal | $N^W, \mathcal{E}^W$ | Node weights, edge weights |
| $S$ | Subject | $M_{\mathcal{T}}^N(\cdot), M_{\mathcal{T}}^{\mathcal{E}}(\cdot)$ | Mapping function from node/edge to type |
| $O$ | Object | $M_W^N(\cdot), M_W^{\mathcal{E}}(\cdot)$ | Mapping function from node/edge to weight |
| $T^{(\cdot)}$ | Triple in $K/K^D/K^H$ | $\tau_{align}^{(\cdot)}$ | Alignment threshold for $E/P$ |
| $m$ | Entity mention | $\mathbb{T}_i$ | Any tree in $XG^{(\cdot)}(q)$ connecting anchors |
| $\mathcal{A}$ | Anchor node | $\mathbb{T}^*$ | GST |

In a nutshell, UNIQORN works as follows. Given an input question, we first retrieve question-relevant *evidences* from one or more knowledge sources using fine-tuned BERT models for sequence pair classification. From these evidences, that are either KG facts or text snippets, UNIQORN constructs a *context graph (XG)* that contains question-specific entities, predicates, types, and candidate answers. Depending upon the input source, this XG thus either consists of: (i) KG facts defining the neighborhood of the question entities, or, (ii) a quasi-KG dynamically built by joining Open IE triples extracted from text snippets, or, (iii) both, in case of retrieval from heterogeneous sources. Triples in the XG originate from evidences that are deemed question-relevant by a fine-tuned BERT model. We identify *anchor nodes* in the XG that match phrases in the question. Treating the anchors as terminals, Group Steiner Trees (GST) are computed that contain candidate answers. These GSTs establish a joint context for disambiguating mentions of entities, predicates, and types in the question. Candidate answers are ranked by simple statistical measures rewarding redundancy within the top-$k$ GSTs. Fig. 1 illustrates this unified approach for the settings of KG-QA, Text-QA, and heterogeneous-QA. UNIQORN thus belongs to the family of methods that locate an answer using approximate graph search and traversal, and does not build an explicit SPARQL(-like) query [99].

**Contributions**. Our salient contributions here are:

- proposing a unified method for answering complex factoid questions over heterogeneous sources comprising KG and text;
- applying Group Steiner Trees as a mechanism for computing answers to complex questions involving multiple entities and relations;
- experimentally comparing UNIQORN on six benchmarks of complex questions against ten state-of-the-art baselines on KGs, text, and heterogeneous sources.

The experiments show that UNIQORN effectively leverages the combination of KG and text, and substantially outperforms all heterogeneous-QA baselines, including the recent UNIK-QA system [89], on complex questions. The gains are particularly massive for the case of zero-shot transfer to questions that are outside the training data. Note that, for UNIQORN, KG is utilized as the primary source (due to its reliability) whereas text is used as a complementary source to balance the incompleteness of knowledge graphs. Temporal staleness of information in these sources is not a concern in this work. Code, data, results, and a demo for this project are available at: `https://uniqorn.mpi-inf.mpg.de`.

**Improvements over QUEST**. UNIQORN is an extension of QUEST [85] for question answering over graphs induced from text, published in SIGIR 2019. UNIQORN contains substantial improvements over QUEST, key factors being:

- QUEST was limited to text inputs, but UNIQORN works over KGs, text, or both.
- UNIQORN incorporates a novel BERT fine-tuning model for scoring the relevance of heterogeneous evidences with respect to the question.
- The evaluation in QUEST solely relied on automated judgements based on two benchmarks. UNIQORN uses large-scale human evaluation with approximately 86$k$ QA pairs judged by Master Workers on Amazon Mechanical Turk.
- QUEST was evaluated on a small dataset of 300 questions. In contrast, experiments for UNIQORN use LC-QuAD 2.0 [41], a large and recent benchmark for complex KG-QA with about 5000 test questions; the Text-QA counterpart was curated by us. Additional experiments with five other benchmarks serve to study the zero-shot transfer performance.
- UNIQORN experiments have ten baselines, whereas QUEST was compared to only three of these.

## 2. Concepts and notation

We now introduce salient concepts necessary for an accurate understanding of UNIQORN. A glossary of concepts and notation is provided in Table 1.

## 2.1. General concepts

**Knowledge graph**. An RDF knowledge graph $K$, like Wikidata or DBpedia, consists of entities $E$ (like *"Leonardo DiCaprio"*), predicates $P$ (like *"award received"*), types $\mathbb{T}$ (like *"film"*), and literals $L$ (like *"07 January 2016"*), organized as a set of subject-predicate-object (SPO) triples $\{T^K\}$ where $S \in E$ and $O \in E \cup \mathbb{T} \cup L$. Optionally, a triple $T_i^K$ may be accompanied by one or more qualifiers as <qualifier-predicate, qualifier-object> tuples that provide additional context. The following is an example of a triple with *qualifiers*[2]: `<LeonardoDiCaprio, awardReceived, AcademyAward; forWork, TheRevenant; pointInTime, 2016>`. Each $T_i^K$ represents a fact in $K$. Under the graph model used in this work, each entity, predicate, type and literal becomes a node in the graph. Edges connect different components of a fact. They run from the subject to the predicate and on to the object of a fact. For facts with qualifiers, there are additional edges from the main predicate to the qualifier predicate and then onwards to the qualifier object. A common alternative model for KGs represents predicates as edge labels, but this makes it difficult to apply our graph algorithms, and to incorporate qualifier information elegantly (considering contextual information contained in qualifiers is a more realistic setting).

**Text corpus**. A text corpus $D$ is a collection of documents, where each document $D_i$ contains a set of natural language snippets $\mathcal{S}$. Each snippet $\mathcal{S}_j$ is defined as a span of text in $D_i$ that contains at least two tokens from the question, within a specified context window. Such documents could come from a static collection like ClueWeb12, Common Crawl, Wikipedia articles, or the open Web. First, we detect all snippets $\{\mathcal{S}_j\}$ in $D_i$ using question token matches inside $D_i$ within the window. Then, to induce structure on $D$, Open IE [32, 5, 87] is performed on each $\mathcal{S}_j \in D_i$, for every $D_i \in D$, to return a set of triples $\{T^{D_i}\}$ in SPO format, where each such triple represents a fact (equivalently, evidence) mined from some $D_i$. These triples are augmented with those built from Hearst patterns [55] run over $D$ that indicate entity-type memberships. This non-canonicalized (open vocabulary) triple store is referred to as a quasi knowledge graph $K^D$ built from the document corpus, where $K^D = \{T^D\} = \bigcup_i T^{D_i}$. Thus, each $T_i^D$ is a fact in $K^D$.

**Evidence**. We refer to a fact in a KG or a snippet from text by the unifying term "evidence".

**Heterogeneous source**. We refer to the mixture of the knowledge graph $K$ and the quasi-KG $K^D$ as a heterogeneous source $K^H = K \cup K^D$, where each triple in this heterogeneous KG can come from either the RDF store or the text corpus.

**Question**. A question $q = \langle q_1 q_2 \ldots \rangle$ is posed either as a full-fledged interrogative sentence (*Who is the director of the western film for which Leonardo DiCaprio won an Oscar?*) or in telegraphic [71, 103] or utterance [1, 138] form (*director of western with Oscar for Leo*), where the $q_i$'s are the tokens in the question. Question tokens are either words, or phrases like entity mentions detected by a named entity recognition (NER) system [95]. Stopwords are not considered as question tokens.

**Complex question**. UNIQORN is motivated by the need for a unified approach to complex questions [111, 85, 122, 41, 114], as simple questions are already well-addressed by prior works [92, 1, 10, 138]. We call a question "simple" if it can be translated into a SPARQL(-like) query or a logical form with a single entity and a single predicate (like *capital of Greece?* $\mapsto$ `Greece capital ?x`). Questions where the simplest proper query requires multiple $E$ or $P$, are considered "complex" (like *director of the western for which Leo won an Oscar?*). There are other notions of complex questions [58], like those requiring grouping and aggregation, (e.g., *which director made the highest number of movies that won an Oscar in any category?*), or when questions involve negations (e.g., *which director has won multiple Oscars but never a Golden Globe?*). These are not considered in this paper.

**Answer**. An answer $a \in A$ to a question $q$ is an entity $e \in E$ or a literal $l \in L$ in the KG $K$ (like the entity `The Revenant` in Wikidata), or a span of natural language text (a sequence of words) from some snippet $\mathcal{S}_j \in D_i$ in the corpus $D$ (like *"The Revenant film"*). $A$ denotes the set of all correct answers to $q$ ($|A| \geq 1$).

## 2.2. Graph concepts

**Context graph**. A context graph $XG$ for a question $q$ is defined as a subgraph of the full / quasi / heterogeneous knowledge graph, i.e., $XG^K(q) \subset K$ (KG) or $XG^D(q) \subset K^D$ (text) or $XG^H(q) \subset K^H$ (mixture of both), such that it contains all triples $T_i^K(q)$ or $T_i^D(q)$ potentially relevant for answering $q$. Thus, $XG(q)$ is expected to contain every answer entity $a \in A$. An $XG$ has nodes $N$ and edges $\mathcal{E}$ with types $(N^\mathcal{T}, \mathcal{E}^\mathcal{T})$ and weights $(N^W, \mathcal{E}^W)$, as discussed below. Thus, an $XG$ is always *question-specific*, and to simplify notation, we often write only $XG$ instead of $XG(q)$. Fig. 1 shows possible context graphs for our running example question in each of the three setups.

**Node types**. A node $n \in N$ in an $XG$ is mapped to one of four categories $N^\mathcal{T}$: (i) entity, (ii) predicate, (iii) type, or (iv) literal, via a mapping function $M_\mathcal{T}^N(\cdot)$, where $n \in E \cup P \cup \mathbb{T} \cup L$. Each $n_j$ is produced from an $S$, $P$, or $O$, from the triples in $K$ or $K^D$. For KG facts, we make no distinction between predicates and qualifier predicates, or objects and qualifier objects. There are no qualifiers in text.

Let us use Fig. 1 for reference. Entities and literals are shown in rectangles with sharp corners, while predicates and types are in rectangles with rounded corners. Even though it is standard practice to treat predicates as edge labels in a KG [112], we model them as nodes, because this design simplifies the application of our graph algorithms.

Note that predicates originating from different triples are assigned unique identifiers in the graph. For instance, for triples $T_1 =$ `<BarackObama, married, MichelleObama>`, and $T_2 =$ `<BillGates, married, MelindaGates>`, we will obtain two `married` nodes, that will be marked as `married-1` and `married-2` in the context graph. Such distinctions prevent false inferences

[2] https://www.wikidata.org/wiki/Help:Qualifiers

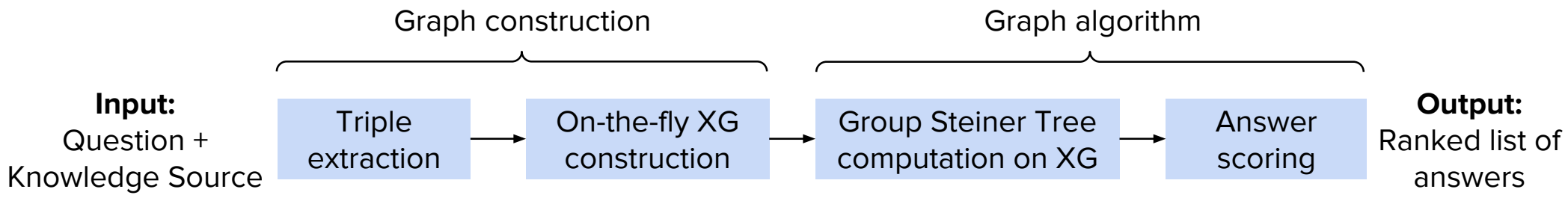

**Figure 2:** System architecture of Uniqorn.

when answering over an XG. For simplicity, we do not show such predicate indexing in Fig. 1.

For text-based $XG^D$, we make no distinction between $E$ and $L$, as Open IE markups that produce these $T^D$ often lack such "literal" annotations. Type nodes $\mathbb{T}$ in $XG^K$ come from the objects of `instanceOf` (for all entities) and `occupation` (for humans only) predicates in $K$ (e.g., for Wikidata), while those in $XG^D$ originate from Hearst patterns [55] in $D$. In Fig. 1a, nodes `TheRevenant`, `director`, `AcademyAwards`, and `2016` are of type $E$, $P$, $\mathbb{T}$, and $L$, respectively. In Fig. 1b, nodes *"The Revenant"*, *"directed by"*, and *"2015 American western film"*, are of type $E$, $P$, and $\mathbb{T}$, respectively.

**Edge types**. An edge $\epsilon \in \mathcal{E}$ in an $XG$ is mapped to one of three categories $\mathcal{E}^{\mathcal{T}}$: (i) triple, (ii) type, or (iii) alignment edge, via a mapping function $M^{\mathcal{E}}_{\mathcal{T}}(\cdot)$. Triples in $\{T^K\}$ or $\{T^D\}$ or $\{T^H\}$, where $M^N_{\mathcal{T}}(O) = \{E \cup L\}$, i.e., the object is of node type entity or literal, contribute triple edges to the $XG$. For example, in Fig. 1a, the two edges between *"The Revenant"* and *"genre"*, and between *"genre"* and *"Western film"*, are triple edges. In Fig. 1b, the two edges between *"Alejandro"* and *"director of"*, and between *"director of"* and *"Survival drama Revenant"*, are triple edges.

Triples where $M^N_{\mathcal{T}}(O) = \{\mathbb{T}\}$ (object is a type) are type triples, and each type triple contributes two type edges to the $XG$. Examples are: edges between `AcademyAwardForBestDirector` and `instanceOf`, and `instanceOf` and `AcademyAwards` in Fig. 1a; and edges between *"The Revenant"* and *"type"*, and *"type"* and *"2015 American western film"* in Fig. 1b.

Alignment edges represent potential synonymy between nodes, and run *only* between nodes of the same type. Alignments are inserted in $XG^D$ or $XG^H$ via external sources of similarity like aliases or word embeddings. There are no alignment edges in $XG^K$ (or more generally, in $K$) as all items in a KG are canonicalized. Examples of alignment edges are the bidirectional dotted edges between *"The Revenant"* and *"Survival drama Revenant"* in Fig. 1b, and `director` and *"directed by"* in Fig. 1c. Insertion of alignment edges as opposed to a naive merging of synonymous nodes is a deliberate choice. This enables more matches with question tokens and a subsequent disambiguation of question concepts. It also precludes the problem of topic drifts from transitive effects of merging nodes into clusters at this stage, and that of choosing a label for such merged clusters.

**Node weights**. A node $n \in N$ in an $XG$ is weighted by a function $M^N_W(\cdot) \in [0, 1]$ according to its similarity to the question. This is obtained by averaging BERT scores of the evidences (Sec. 3.2) of which $n$ is a part. Node weights are used as a potential criterion for answer ranking.

**Edge weights**. Each edge $\epsilon = (n_i, n_j)$ in an $XG$ is assigned a weight by a function $M^{\mathcal{E}}_W(\cdot) \in [0, 1]$. Edge weights are later converted to edge costs that are vital for computing Group Steiner Trees on the XG. In Uniqorn, edge weights are the BERT scores of evidences (KG facts or text snippets) from where the edge originates (Sec. 3.2). Edge weights are averaged if the same edge is found in multiple evidences. This holds for triple and type edges.

Weights of alignment edges come from similarities between nodes in the XG. For entities, this is computed as the Jaccard overlap of character-level trigrams [138] between the node labels that are the endpoints of the edge (for KGs, entity aliases available as part of the KG are appended to entity names before computing the similarity). Character-level $n$-grams make the matching robust to spelling errors. Lexical matching is preferred for entities and literals as we are more interested in *hard equivalence* rather than a soft relatedness. For predicates and types, lexical matches are not enough and semantic similarity computations are necessary. So alignment scores are computed as pairwise embedding similarities (cosine values) over words in the two node labels (one word from each node), followed by a maximum over these pairs. This is then min-max normalized to $[0, +1]$ from $[-1, +1]$. Wikipedia2vec [134], that taps into both corpus statistics and link structure in Wikipedia, was used for computing predicate and type embeddings.

An alignment edge is inserted into an $XG^D$ or $XG^H$ if the similarity exceeds or equals some threshold $\tau_{align}$, i.e., $\text{sim}(\text{label}(n_i), \text{label}(n_j)) \geq \tau_{align} \in (0, 1]$. Zero is not an acceptable value for $\tau_{align}$ as that would mean inserting an edge between every pair of possibly unrelated nodes. This alignment insertion threshold $\tau_{align}$ could be potentially different for entities ($\tau^E_{align}$) and predicates ($\tau^P_{align}$), due to the use of different similarity functions. These (and other) hyper-parameters are tuned on a dev set of QA pairs.

**Anchor nodes**. A node $n$ in an $XG$ is an anchor node $\mathcal{A}$ if it matches one of the question tokens. Such matches may either be lexical (*"western"* $\mapsto$ `2015 American western film` in Fig. 1b), or more sophisticated mapping of entity mentions $m$ in questions to KG entities $\{e\}$ via named entity recognition and disambiguation (NERD) systems [80, 57, 44] (*"Leo"* $\mapsto$ `Leonardo DiCaprio` in Fig. 1a). Anchors are grouped into sets, where a set $\mathcal{A}^i$ is defined as $\{\mathcal{A}^i_1, \mathcal{A}^i_2, \dots\}$, depending upon which question token $q_i$ the elements of the set match. In other words, more than one XG node can match the same question token, and hence the need for the superscript $i$: the anchor nodes corresponding to $q_2$ would be denoted by $\{\mathcal{A}^2_1, \mathcal{A}^2_2, \dots\}$. For example, *"director"* in the question matches nodes *"director of"*, *"directed by"*, *"Best Director"*, and so on, in Fig. 1b. Anchors thus identify the question-relevant nodes of the $XG$, in the vicinity of which answers can be expected to be found. Any node of category entity, predicate, type, or literal, can qualify as an anchor.

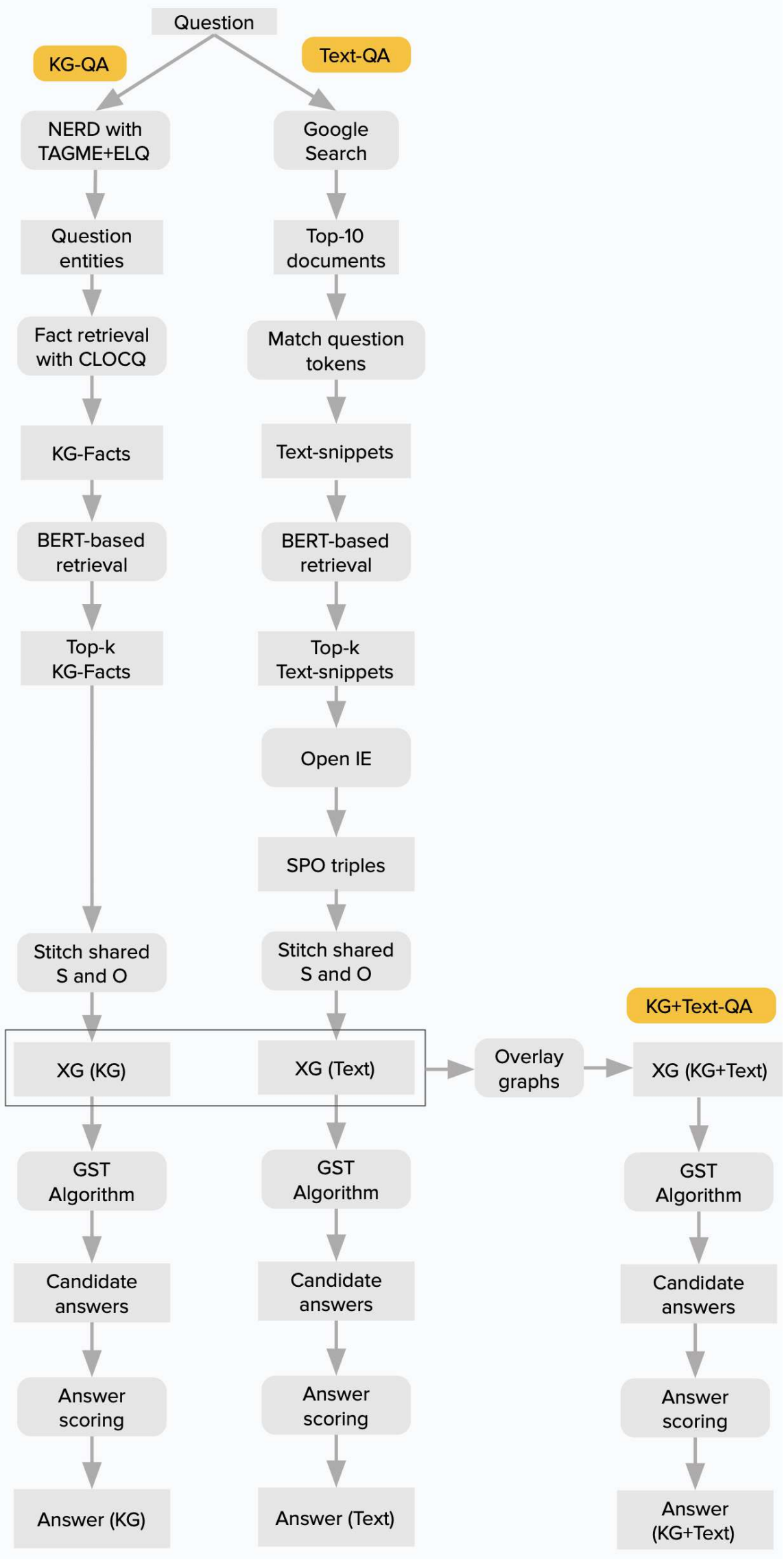

**Figure 3:** UNIQORN workflows for the three answering setups.

## 3. UNIQORN: Graph construction

Fig. 2 gives an overview of the UNIQORN architecture. The two main stages – construction of the question-specific context graph (Sec. 3), and computing answers by Group Steiner Trees (Sec. 4) on this context graph – are described in this section and the next. Individual workflows for the KG, Text, and KG+Text setups are illustrated in Fig. 3.

We describe the XG construction, using our example question for illustration: *director of the western for which Leo won an Oscar?* Instantiations of key factors in the two settings, KG-QA and Text-QA, are shown in Table 2. The heterogeneous setting can be understood as simply the union of the two individual cases. The context graph is built in two stages: (i) identifying question-relevant evidences from the knowledge sources, and (ii) creating a graph from these question-relevant evidences.

### 3.1. Retrieving question-relevant evidences

Our goal is to reduce huge knowledge repositories to reasonably-sized question-relevant subgraphs, over which graph algorithms can be run with interactive response times. **From knowledge graph**. A typical curated KG $K$ contains billions of facts, with millions of entities and thousands of predicates, occupying multiple terabytes of disk space. To reduce this enormous search space in $K$, we first identify entities $E(q)$ in the question $q$ by using methods for named entity recognition and disambiguation (NERD) [44, 57, 63, 80]. Specifically, to link mentions of entities in the question to knowledge graph entities, we use the named entity recognition and disambiguation tools TAGME [44] and ELQ [80]. These tools return Wikipedia URLs that we can easily map to Wikidata entities, using the linking information already present in Wikidata. This produces KG entities (`LeonardoDiCaprio`, `AcademyAwards`) as output. Next, all KG triples are fetched that are in the 1-hop neighborhood of an entity in $E(q)$[3]. To reduce this large and noisy set of facts (equivalently, KG-evidences) to a few question-relevant ones (to obtain $T^K(q)$), we fine-tune BERT [34] (Sec. 3.2).
**From text corpus**. The Web is the analogue of large KGs in Text-QA. Similar to the case for the KG, we collect a small set of potentially relevant documents for $q$ from $D$ using Google Search, with $D$ being the Web. Alternatively, one could use an IR system like ElasticSearch when $D$ is a fixed corpus such as Wikipedia.

KG-style entity-centric retrieval is not a practical approach for open-domain text: it requires entity linking on a potentially enormous set of sentences, highly limiting the efficiency of an on-the-fly procedure. Rather, we take a noisy and recall-oriented approach: we locate question tokens (stopwords are not considered) within the relevant documents, and consider a window of words (window length = 50 in all our experiments) to each side of a match as a question-relevant snippet $S$. In case two snippets overlap, they are merged to form a longer snippet. These snippets form our candidate text-evidences for locating an answer, and analogous to the KG facts obtained via NERD, are passed on to a BERT-based pruning model.

### 3.2. Finding top evidences with BERT

**Training a classifier for question-relevance**. The goal of our fine-tuned BERT model is to classify an evidence as being relevant to the question or not. In other words, we want a model that can score each evidence based on its likelihood of containing the answer. To build such a model, we prepare training data as follows. The KG fact/text snippet retrieval step above feeds a set of evidences to the classifier. Out of these evidences, the ones that contain the gold answer(s) are labeled as *positive instances*. This is a *distant or noisy supervision signal* as the mere *presence* of an answer is not necessarily conclusive evidence of relevance to the question: the fact contained in the evidence may not be the one intended in the question. As an example, consider the question *Who played Hugh in The Revenant?*: the KG fact ⟨`The Revenant, award received, Academy Award for Best Actor; winner, Leonardo DiCaprio`⟩ would be deemed relevant due to the presence of the gold answer `Leonardo DiCaprio` in it. Nevertheless, the method works: such false positives are rare in reality. For each positive instance, we randomly sample five *negative instances* from the evidences that do not contain the answer. Sampling question-specific negative instances helps learn a discriminative classifier, as all negative instances are guaranteed to contain at least one entity/keyword from the question. Using *all facts* that do not contain an answer as negative examples would result in severe class imbalance and efficiency bottlenecks during training.

[3]Moving from one entity to another entity on the knowledge *graph* is considered one "hop". The 2-hop entity neighborhood in a KG can be enormously large, especially for popular entities like countries (`UK`) and football clubs (`FC Barcelona`), with thousands of 1-hop neighbors. This problem is exacerbated by proliferations via type nodes (all humans are within two hops of each other on Wikidata). See [23] for more discussion on KG neighborhoods.

**Table 2**
Instantiations of different factors of XGs from KG and text corpus.

| Scenario | KG | Text |
|---|---|---|
| Triples in XG | NERD + KG-lookups + BERT | IR system + BERT + Open IE |
| Node types | Entity, predicate, type, literal | Entity, predicate, type |
| Type nodes | `instance of`, `occupation` triples | Hearst patterns |
| Edge types | Triple, alignment, type | Triple, alignment, type |
| Entity alignments | None | With character trigrams |
| Predicate and type alignments | None | With word embeddings |
| Node weights | With BERT | With BERT |
| Edge weights | With BERT | With BERT |

We then pool together the positive and negative ⟨question, evidence⟩ pairs for all training questions. Text snippet evidences are already in natural language and amenable to be passed directly through a BERT encoder. To bring KG fact evidences closer to an NL form, we *verbalize* them by concatenating their constituents; qualifier statements are joined using *"and"* [89]. For example, the KG-fact ⟨`The Revenant, nominated for, Academy Award for Best Director; nominee, Alejandro González Iñárritu`⟩ with one qualifier would be verbalized as *"The Revenant nominated for Academy Award for Best Director and nominee Alejandro González Iñárritu"*. The questions and the verbalized evidences, along with binary ground-truth labels, are fed as training input to a *sequence pair classification model* for BERT: one sequence is the question, the other is the evidence.

**Applying the classifier**. Following [34], the question and the evidence texts are concatenated with the special separator token `[SEP]` in between, and the special classification token `[CLS]` is prepended to this sequence. The final hidden vector corresponding to `[CLS]`, denoted by $C \in \mathbb{R}^H$ ($H$ is the size of the hidden state), is considered to be the accumulated representation. Weights $W$ of a final classification layer are the only new parameters introduced during fine-tuning, where $W \in \mathbb{R}^{\mathcal{L} \times H}$, where $\mathcal{L}$ is the number of class labels ($\mathcal{L} = 2$ here, as an evidence is either question-relevant or it is not). The value $\log(\text{softmax}(CW^T))$ is used as the classifier loss. Once this classifier is trained, given a new ⟨question, evidence⟩ pair, it outputs the probability (and the label) of the evidence containing an answer to the question. We make this prediction for all candidate evidences retrieved for a question, and sort them in descending order of this question relevance likelihood. We pick the top-$\varepsilon$ evidences from here as our question-relevant set for constructing the XG.

## 3.3. Materializing the context graph

**For KG evidences**. Evidences from the KG are facts, and can be trivially cast into a graph that is a much smaller subgraph of the entire KG. Entities, predicates, types, and literals constitute nodes, while edges represent connections between pieces of the same fact (recall Fig. 1a). Each distinct entity, type, and literal maps to one node. For predicates, however, each possible repetition gets its own node (recall the risk of false inference via predicate instance merging from Sec. 2.2). We then add type information to this graph, useful for QA in several ways [99, 3, 137, 142]. For this, we look up the KG types for each entity in the qualifying set of evidences, and additionally look up occupations for humans (recall Sec. 2.2), and add these triples to the graph. To ensure connectivity in this graph, as far as possible, we add shortest paths from the KG between NERD entities detected in this question to the context graph (complex questions often have more than one entity). This step usually helps reintroduce question-relevant connections in the context graph between entities in the question. Such shortest paths can be obtained via the CLOCQ[4] API. The largest connected component (LCC) is extracted from this graph, and the final structure thus obtained constitute $XG^K(q)$, being made up of individual triples $T^K(q)$. The LCC is expected to cover the most pertinent information with respect to the input question.

**For text evidences**. There is no natural graph structure in the NL snippets, so we induce it using a simple version of open information extraction (open IE). The goal of open IE is to extract informative triples from raw text sources. As off-the-shelf tools like Stanford OpenIE [5], OpenIE 5.1 [101], ClausIE [32], or MinIE [51], all have limitations regarding either precision or recall or efficiency, we developed our own custom-made open IE extractor. We start with part-of-speech (POS) tagging and named entity recognition (NER) on the original sentences from $D$ (and not on the snippets $\mathcal{S}$, to preserve necessary context information vital for such taggers). This is followed by lightweight coreference resolution by replacing each third person personal and possessive pronoun (*"he, him, his, she, her, hers"*) by its nearest preceding entity of type PERSON. We then define a set of POS patterns that may indicate an entity or a predicate. Entities are marked by an unbroken sequence of nouns, adjectives, cardinal numbers, or mention phrases from the NER system (e.g., *"Leonardo DiCaprio"*, *"2016 American western film"*, *"Wolf of Wall Street"*). To capture both verb- and noun-mediated relations [133], predicates correspond to POS patterns verb, verb+preposition, or noun+preposition (e.g., *"produced"*, *"collaborated with"*, *"director of"*). See node labels in Fig. 1b for examples. The patterns are applied to each snippet $\mathcal{S}_j \in D$, to produce a markup like $\mathcal{S}_j \equiv E_1 \dots E_2 \dots P_1 \dots E_3 \dots P_2 \dots E_4$. The ellipses ($\dots$) denote intervening words in $\mathcal{S}_j$. From this markup, Uniqorn finds all $(E_{i_1}, E_{i_2})$ pairs that have exactly one predicate $P_k$ between them, this way creating triples $\langle E_1, P_1, E_3 \rangle$ and $\langle E_2, P_1, E_3 \rangle$ (but excluding $\langle E_1, P_2, E_4 \rangle$). Patterns from [133], specially designed for noun phrases as relations (e.g., *"Oscar winner"*), are applied as well. Snippets that contain two (or more) entities but no intervening predicate contribute triples with a special cooccurs predicate. For example, if a snippet contains three entities $E_1, E_2, E_3$ but zero predicates, we would add triples: $\langle E_1$, cooccurs, $E_2\rangle$, $\langle E_1$, cooccurs, $E_3\rangle$, $\langle E_2$, cooccurs, $E_3\rangle$ to the open IE triples. This rule helps tap into information in snippets like *"Leonardo was in Inception"*, where the intended relation is implicit (*"starred"*). The rationale for our heuristic extractor is to achieve high answer *recall*, at the cost of introducing noise. The noise is handled in the answering stage later.

[4] `https://clocq.mpi-inf.mpg.de/`

Just like for KG evidences, we would like to extract entity type information for text as well. To this end, we leverage Hearst patterns [55], like *"... $NP_2$ such as $NP_1$ ..."* (matched by, say, *"western films such as The Revenant"*) *"... $NP_1$ is a(n) $NP_2$ ..."* (matched by, say, *"The Revenant is a 2015 American western film"*), or *"... $NP_1$ and several other $NP_2$ ..."* (matched by, e.g., *"Alejandro Iñàrritu and many other Mexican film directors"*). Here $NP$ denotes a noun phrase, detected by a constituency parser[5]. The resulting triples about type membership (of the form: $\langle NP_1$, type, $NP_2\rangle$ for e.g., $\langle$The Revenant, type, 2015 American western film$\rangle$) are added to the triple collection. Finally, as for the KG-QA case, all triples are joined by $S$ or $O$ with exact string matches, and the LCC is computed, to produce the final $XG^D(q)$. To compensate for the diversity of surface forms where different strings may denote the same entity or predicate, alignment edges are inserted into $XG^D(q)$ for node pairs as discussed in Sec. 2.2. $XG^D(q)$ is made up of individual triples $T^D(q)$. Thus, a large quasi KG $K^D$ is never materialized, and we directly construct $XG^D$, on which our graph algorithms are run.

[5] https://stanfordnlp.github.io/CoreNLP/parse.html

**For heterogeneous evidences**. The union of $XG^K$ and $XG^D$ form $XG^H$, the final heterogeneous context graph for question $q$. A BERT model was fine-tuned on the mixture of KG and text repositories, the top-$\epsilon$ evidences were retrieved from each source, and processed as described above. The pool of resultant triples from both sources comprise the heterogeneous context graph $XG^H$.

For a quick recap of how Uniqorn works through the three setups of QA over KG, Text, and KG+Text, the reader is referred once more to Fig. 3.

## 4. Uniqorn: Graph Algorithm

For the given context graph $XG(q)$, we find candidate answers $a \in A$ as follows. First, nodes in the $XG$ are identified that best capture the question; these nodes become anchors. For $XG^K$, question entities detected by NERD systems become entity anchors. For predicates, types, and literals, any node with a token in its label that matches any of the question tokens, becomes an anchor (cast member becomes an anchor if the question has *Who was cast as ...?*). To ensure better semantic coverage, node labels are augmented with aliases from the KG, that are a rich yet underexplored source of paraphrases. For example, Wikidata contains the following aliases for cast member: *"starring"*, *"actor"*, *"actress"*, *"featuring"*, and so on. Thus, cast member will become an anchor node if the question has any of the synonyms above. For $XG^D$, any node with a token in its label that matches any of the question tokens becomes an anchor. Node and edge weights in $XG^K$ or $XG^D$ are obtained by using the BERT-based scores that the original evidence (source of the node or edge) received from the fine-tuned model. If a node or edge originates from multiple evidences, their scores are averaged. This helps us harness *redundancy* of information across evidences.

Anchors are grouped into equivalence classes $\{\mathcal{A}^k\}$ based on the question token that they correspond to. At this stage, we have the directed and weighted context graph $XG$ as a 6-tuple: $XG^{(\cdot)}(q) = (N, \mathcal{E}, N^T, \mathcal{E}^T, N^W, \mathcal{E}^W)$. For simplicity, we disregard the direction of edges, turning $XG$ into an undirected graph.

**Group Steiner Tree**. We postulate that the criteria for identifying good answer candidates in the XG are as follows: (i) answers lie on *paths* connecting anchors; (ii) *shorter* paths with *higher* weights are more likely to contain correct answers; and (iii) including at least one instance of an anchor from each *group* is necessary, to satisfy all conditions in a complex question $q$. Formalizing these desiderata leads us to the notion of *Steiner Trees* [42, 13, 72, 75]: for an undirected weighted graph, and a subset of nodes called *"terminals"*, find the tree of least cost that connects all terminals. If the number of terminals is two, then this is the weighted shortest path problem, and if all nodes of the graph are terminals, this becomes the minimum spanning tree (MST) problem. In our setting, the graph is the $XG$, and terminals are the anchor nodes. As anchors are arranged into groups, we pursue the generalized notion of *Group Steiner Trees*

*(GST)* [50, 81, 39, 109, 18, 113]: compute a minimum-cost Steiner Tree that connects at least one terminal from each group, where weights of edges $\mathcal{E}_i$ are converted into *costs* as $\text{cost}(\mathcal{E}_i) = 1 - M_W^{\mathcal{E}}(\mathcal{E}_i)$. At this point, the reader is referred to Fig. 1 again, for illustrations of what GSTs look like (shown in orange). To tackle questions with a chain-join [99] component (a chain join question contains an indirection, like *profession of father of DiCaprio?*), the complete evidences from where the predicate anchors are extracted, are admitted into the GST. Without this step, the predicate nodes can be imagined as somewhat "dangling" from the Group Steiner tree. We detect chain join components with simple patterns like the presence of more than one *"of"* in the question, or a combination of *"of"*s and apostrophes.

Formally, the GST problem in our setting can be defined as follows. Given a question $q$ with $|q|$ tokens, an undirected and weighted graph $XG(q) = (N, \mathcal{E})$, and groups of anchor nodes $\{\mathcal{A}^1, \dots \mathcal{A}^{|q|}\}$ with each $\mathcal{A}^i \subset N$, find the minimum-cost tree $\mathbb{T}^* = (N^*, \mathcal{E}^*) = \arg\min_j \text{cost}(\mathbb{T}_j, XG)$, where $\mathbb{T}_j$ is any tree that connects at least one node from each of $\{\mathcal{A}^1, \dots \mathcal{A}^{|q|}\}$, such that $\mathcal{A}^i \cap N^* \neq \phi$ for each $\mathcal{A}^i$, and cost of $\mathbb{T}_j$, $\text{cost}(\mathbb{T}_j) = \sum_l \text{cost}(\mathcal{E}_l)$, where $\mathcal{E}_l \in \mathbb{T}_j$.

While finding the GST is an NP-hard problem, there are approximation algorithms [50], and also exact methods that are fixed-parameter tractable with respect to the number of terminal nodes. Since approximation may be detrimental to factoid QA that needs precise answers, we adapted the method of Ding et al. [39] from the latter family, which is exponential in the number of terminals but $O(n \ log \ n)$ in the graph size $n$. Luckily for us, the number of terminals (anchors) is indeed typically much less of a bottleneck than the sizes of the $XG$ in terms of nodes ($n$). Specifically, the terminals are the anchor nodes derived from the question tokens – so their numbers are not prohibitive with respect to computational costs (a question is generally not very long). Actual runtimes are presented in Sec. 6.2.

The algorithm is based on *dynamic programming* and works as follows. It starts from a set of singleton trees, one for each terminal group, rooted at one of the corresponding anchor nodes. These *trees are grown* iteratively by exploring immediate neighborhoods for least-cost nodes as expansion points. Trees are merged when common nodes are encountered while growing. The trees are stored in an efficient implementation of priority queues with *Fibonacci heaps* [47]. The process terminates when a GST is found that covers all terminals (i.e., one per group). Bottom-up dynamic programming guarantees that the result is *optimal*.

**Relaxation to top-$k$ GSTs**. It is possible that the GST simply connects terminals from each anchor group directly, without having any internal nodes at all, or with only predicates and/or types as internal nodes. Since we need entities or literals as answers, such possibilities necessitate a relaxation of our solution to compute a number of top-$k$ least-cost GSTs. GST-$k$ ensures that we always get a non-empty answer set, albeit at the cost of making some detours in the graph. Moreover, using GST-$k$ provides a natural answer ranking strategy, where the score for an answer can be appropriately reinforced if it appears in multiple such low-cost GSTs. This postulate, and the effect of $k$, is later quantified in our experiments. Note that since the tree with the least cost is always kept at the top of the priority queue, the $k$ trees can be found in the increasing order of cost, and no additional sorting is needed. In other words, the priority-queue-based implementation of the GST algorithm [39] automatically supports this top-$k$ computation. The time and space complexities for obtaining GST-$k$ is the same as that for GST-1. Fig. 4 gives an example of GST-$k$ ($k = 3$).

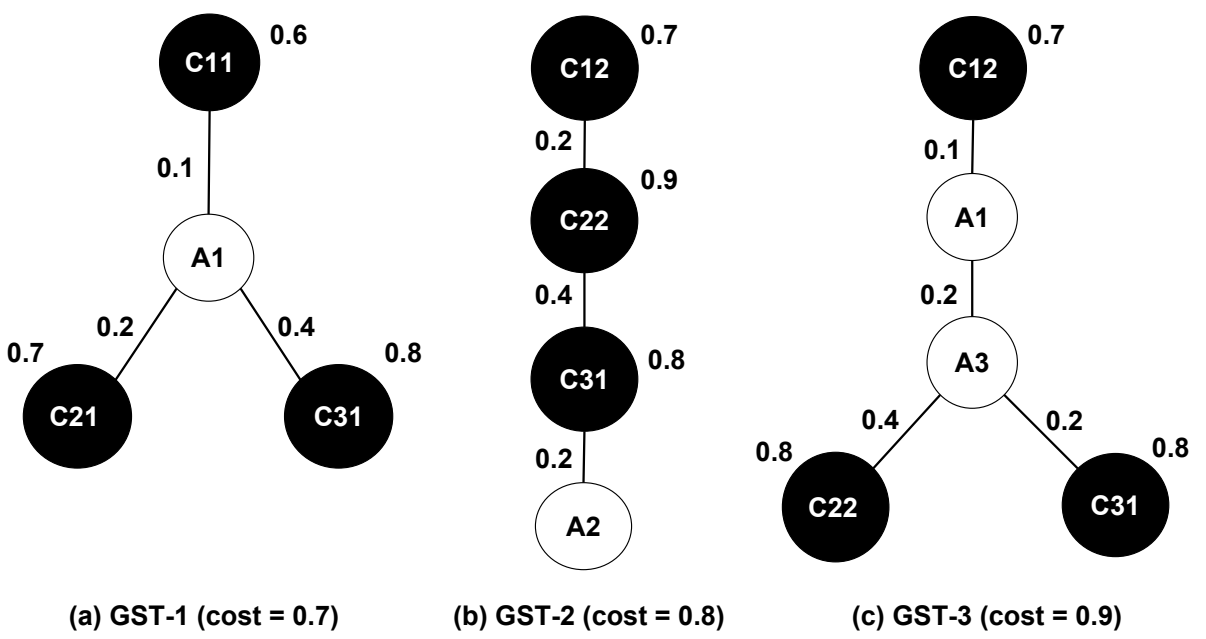

**Figure 4:** Illustrating GST-$k$, showing edge costs and node weights. Anchors (terminals C11, C12, ...) and answer candidates (non-terminals A1, A2, A3) are shown in black and white circles respectively. {(C11, C12), (C21, C22), (C31)} represent anchor groups. Edge costs are used in finding GST-$k$, while node weights may be used in answer ranking. Situation (b) is likely to arise from chain-join support. A1 is likely to be a better answer due to its presence in two GSTs in the top-3.

**Answer ranking**. Non-terminal entities in GSTs are candidates for final answers. However, this mandates ranking. We use the number of top-$k$ GSTs that an answer candidate lies on, as the ranking criterion. Alternatives like weighting these trees by their total node weight, tree cost, or the answers' proximity to anchor nodes, are investigated in Sec. 6.2. The top-ranking answer is presented to the end user.

## 5. Experimental setup

This section explains the experimental setup that we use to evaluate our method Uniqorn, and compare it with state-of-the-art. We organize this large section into benchmarks, systems tested, metrics, and human evaluation.

### 5.1. Benchmarks

We first explain the six benchmarks used in this work, and the associated knowledge sources and QA pairs.

#### *5.1.1. Knowledge sources*

As our **knowledge graph** we use the NTriples dump of the full Wikidata[6] as of 31 January 2022, including all qualifiers. The original dump consumed about 2 TB of disk space, and contained about 12B triples. We use a cleaned version[7] that prunes language tags, external identifiers, additional schema labels, and so on (KB cleaning steps in [23]). This left us with about 2B triples with $\simeq$ 40 GB disk space. The cleaned KB is accessed via the recently proposed Clocq API[8]. Note that there is nothing in our method specific to Wikidata, and can be easily extended to other KGs like YAGO [110] or DBpedia [7]. Wikidata was chosen as it is one of the popular choices in the community today, and has an active community that contributes to growth of the KG, akin to Wikipedia (both are supported by the Wikimedia foundation). Further, in 2016, Google ported a large volume of the erstwhile Freebase into Wikidata [90].

[6] https://dumps.wikimedia.org/wikidatawiki/entities/
[7] https://github.com/PhilippChr/wikidata-core-for-QA

**Table 3**
Test sets with sampled questions from each benchmark, totaling about $7k$ complex questions.

| Benchmark | #Questions | Answers |
|---|---|---|
| Complex questions from LC-QuAD 2.0 [41] | $4,921$ | Wikidata entities |
| Complex questions from LC-QuAD 1.0 [119] | $1,459$ | DBpedia entities, mapped to Wikidata via Wikipedia |
| Complex questions from WikiAnswers (CQ-W) [3] | 150 | Freebase entities, mapped to Wikidata via Wikipedia |
| Complex questions from Google Trends (CQ-T) [85] | 150 | Entity mention text, manually mapped to Wikidata |
| Complex questions from QALD $1-9$ [121] | 70 | DBpedia entities, mapped to Wikidata via Wikipedia |
| Complex questions from ComQA [2] | 202 | Wikipedia URLs, mapped to Wikidata entities |
| Total number of complex questions | $6,952$ | Wikidata entities |

For **text**, we create a pool of ten Web pages per question, obtained from Google Web search in January 2022. We issued the full question as a query to the Google Custom Search API[9], and create a question-specific corpus from these top-10 results (full Web pages, not just snippets).

In the API, we send a 'query' request[10]. At a time, links to ten Web pages get crawled, and then we scrape the contents of these pages using the `urllib.request` library[11]. If we do not get contents of the top-10 pages, we try to get replacements from the API up to four times with 'start' indices = 1, 11, 21 and 31 (equivalent to going to the fourth search result page) and stop whenever we are able to scrape the contents of ten Web pages. In "Search Settings", we used the following configuration: "Image Search": "OFF", "SafeSearch": "OFF", "Augment results": "ON", "Search the entire Web": "ON", "Region": "All Regions", "Sites to Search": " *.google.com/* ", and "Language": "English".

This design choice of fetching pages from the Web was made to be close to the direct-answering-over-the-Web-setting, and not be restricted to specific choices of benchmarks that have associated corpora. This also enables comparing baselines from different QA families on a fair testbed. This was done despite the fact that it entailed significant resources at our end, as this had to be done for thousands of questions. *In addition*, text in these Web documents were entity-linked to Wikidata to enable training some of the supervised baselines. For reproducibility, the question-specific text collections (ten complete documents per question with HTML tags pruned with Beautiful Soup[12], and not just snippets) are made available at `https://uniqorn.mpi-inf.mpg.de`, along with relevant code and results.

The **heterogeneous** answering setup was created by considering the above two sources together. In other words, each question from our benchmark is answered over the *combination* of the entire KG *and* the corresponding question-specific text corpus.

*All baselines were exposed to the same knowledge sources as* Uniqorn, in all setups – KG, Text, and KG+Text.

### *5.1.2. Question-answer pairs*

There is no QA benchmark that is directly suitable for answering *complex* questions over *heterogeneous* sources. ComplexWebQuestions [114] comes the closest, but it is somewhat outdated in today's landscape, given that it has only search snippets instead of complete documents, making it unsuitable for evaluating Text-QA models that can handle bigger contexts. Moreover, it relies on the Freebase KG that is deprecated since 2016. As a result, we choose six relatively recent benchmarks (Table 3) with realistic questions, proposed for KG-QA, and curate text corpora for questions in these benchmarks. Note that the opposite strategy of adapting Text-QA benchmarks for KG and heterogeneous QA, is not a practical option. This is driven by the rationale that answers in several mainstream Text-QA benchmarks like SQuAD [97], TriviaQA [70], HotpotQA [136], NaturalQuestions [76], or WikiQA [135], may be arbitrary spans or sentences of text from given passages, and may not map to crisp *entities* – and hence would be out of scope for this work on factual QA. Further, many of these benchmarks do not have substantial volumes of *complex questions*. So out of these six chosen datasets, the most recent one, LC-QuAD 2.0, served as the main benchmark (owing to its relatively larger size), and the rest were used to test the generalizability of QA systems in zero-shot experiments (running pre-trained models on unseen questions). Baselines that require supervision were trained on the full original LC-QuAD 2.0 train set (18k questions), and hyperparameter tuning for all methods were done on a random sample of 1000 questions from the LC-QuAD 2.0 development set (6k questions). Uniqorn only requires tuning of the hyperparameters inside the BERT

[8] `https://clocq.mpi-inf.mpg.de/`

[9] `https://developers.google.com/custom-search/v1/overview`

[10] `https://www.googleapis.com/customsearch/v1?key=<INSERT-YOUR-KEY>&cx=<INSERT-YOUR-CX>&q='+query+'&start='+str(start)`

[11] `https://docs.python.org/3/library/urllib.request.html`

[12] `https://code.launchpad.net/beautifulsoup/`

**Table 4**
Examples of complex questions from each benchmark.

| Benchmark | Complex Questions |
|---|---|
| LC-QuAD 2.0 [41] | *Which of Danny Elfman's works was nominated for an Academy Award for Best Original Score?* |
| | *When did Glen Campbell receive a Grammy Hall of Fame award?* |
| LC-QuAD 1.0 [119] | *Which award that has been given to James F Obrien, had used Laemmle Theatres for some service?* |
| | *Which home stadium of 2011-12 FC Spartak Moscow season is also the location of birth of Svetlana Gounkina?* |
| CQ-W [3] | *Which actor is married to Kaaren Verne and played in Casablanca?* |
| | *Who graduated from Duke University and was the president of US?* |
| CQ-T [85] | *In which event did Taylor Swift and Joe Alwyn appear together?* |
| | *Who played for Barcelona and managed Real Madrid?* |
| QALD $1-9$ [121] | *Which street basketball player was diagnosed with Sarcoidosis?* |
| | *Which recipients of the Victoria Cross fought in the Battle of Arnhem?* |
| ComQA [2] | *What is the river that borders Mexico and Texas?* |
| | *Who killed Dr Martin Luther King Jr's mother?* |

model and the alignment thresholds, and for this, used only the 1000-question dev sample above.

To ensure that the benchmark questions pose sufficient difficulty to the systems under test, all questions from individual sources were manually examined to ensure that they have at least two entity mentions *or* at least two relations. Questions that do not have a ground truth answer in our Wikidata KG are dropped. Aggregations [127] (questions with numerical counts as answers [52]), and Boolean existential questions [100] (questions with yes/no answers) were also removed, as these are out of scope of this work. The number of questions finally contributed by each source to our evaluation is shown in Table 3. These are factoid questions: each question usually has one or a few entities as correct answers (72% of the total number of questions across all benchmarks have exactly one ground truth answer, and most of the rest have $2-3$ gold answers). Details of specific benchmarks are provided below. Two examples of complex questions from each benchmark are shown in Table 4.

(i) **LC-QuAD 2.0** [41]: This very large QA benchmark was compiled using crowdsourcing with human annotators verbalizing KG templates that were shown to them. Answers are Wikidata/DBpedia SPARQL queries that can be executed over the corresponding KGs to obtain entities/literals as answers. This serves as our main benchmark for all experiments and analysis.

(ii) **LC-QuAD 1.0** [119]: This dataset was created by a similar process as LC-QuAD 2.0, but where the crowdworkers directly corrected an automatically generated natural language question. Answers are DBpedia entities – we linked them to Wikidata for our use via their Wikipedia URLs, that act as bridges between popular curated KGs like Wikidata, DBpedia, Freebase, and YAGO.

(iii) **CQ-W** [3]: These questions were sampled from the WikiAnswers community QA platform (Complex Questions from WikiAnswers, hence CQ-W). Answers are Freebase entities. We mapped them to Wikidata using Wikipedia.

**Table 5**
Basic properties of UNIQORN's XGs, averaged over LC-QuAD 2.0 test questions. Alignment edge thresholds affect these sizes: for each setup, the dev-set-tuned optimal-value for answering performance was used in these measurements. The graphs typically get denser as the alignment edge insertion thresholds are lowered. Values in the "KG+Text" row may not be exact additions of the two settings due the computation of the largest connected component for each setting independently, and slight parameter variations. Expectedly, the heterogeneous setting has the largest number of nodes and edges.

| Setup | #Nodes | | | #Edges | | |
|---|---|---|---|---|---|---|
| | Entity | Predicate | Type | Triple | Alignment | Type |
| **KG+Text** | 215.2 | 255.6 | 12.4 | 508.6 | 6,151.9 | 2.5 |
| **KG** | 10.3 | 27.6 | 11.5 | 38.7 | N. A. | 16.4 |
| **Text** | 147.9 | 182.3 | 3.6 | 362.6 | 3.9 | 4.3 |

(iv) **CQ-T** [85]: This dataset contains complex questions about emerging entities created from queries in Google Trends (Complex Questions from Google Trends, hence CQ-T). Answers are text mentions of entities, that we manually map to Wikidata.

(v) **QALD** [121, 84]: We collated these questions by going through nine years of QA datasets from the QALD benchmarking campaigns (editions $1-9$). Answers are DBpedia entities, that we map to Wikidata using Wikipedia.

(vi) **ComQA** [2]: These factoid questions were also sampled from the WikiAnswers community QA corpus. Answers are Wikipedia URLs or free text, which we map to Wikidata.

## 5.2. Systems under test

We now detail the configuration of UNIQORN, and list the ten baselines that we used for comparison.

### *5.2.1.* UNIQORN *configuration*

As mentioned in Sec. 3.1, we need NERD systems for the KG-QA pipeline. To improve answer recall, we used two complementary NERD systems on questions – one biased towards precision (Elq [80], with default configuration), and one towards recall (Tagme [44], with zero cut-off threshold $\rho$). Both systems disambiguate to Wikipedia, and we map them to Wikidata using links available in the KG. All baselines were also given the advantage of these same disambiguations. For Text-QA, the question was issued to Google Web Search and the top-10 documents returned served as the initial corpus. POS tagging and NER on questions and these retrieved documents were done using spaCy[13]. For simplicity, POS tags were mapped to the Google Universal Tagset [93]. These POS tags were used for running our open IE pattern extractors, and identifying noun phrases necessary for matching Hearst patterns. Entity alignment edges were inserted using Jaccard coefficients on KG item labels concatenated with aliases (character trigrams used for constructing the sets). Predicate and type alignments were added using cosine similarities between 100-dimensional Wikipedia2vec [134] embeddings of the respective pairs of nodes (Sec 2.2). BERT fine-tuning needs a train-dev split (the best model is then applied on the test set): the 1000-question LC-QuAD 2.0 development set was split in an 80:20 ratio for this purpose. The following hyperparameters were found to work best on the development set: a batch size of 50, a learning rate of $3 \times 10^{-5}$, and a gradient accumulation of 4. The maximum token length was 512. Hugging Face libraries for the BERT-base-cased model[14] was used, for the sequence pair classification task[15]. The top-5 evidences were returned from the BERT fine-tuning, i.e. $\epsilon = 5$. The three hyperparameters for the GST step ($\tau^E_{align}$, $\tau^P_{align}$, and top-$k$ GSTs to consider) were tuned on the 1000-question development test using grid search in each of the three settings. We obtained: (i) for KG+text: $\tau^E_{align} = 0.8, \tau^P_{align} = 0.7, k = 10$; (ii) for KG: $k = 10$ (no alignment necessary for KGs); and (iii) for text: $\tau^E_{align} = 0.5, \tau^P_{align} = 0.9, k = 10$. Using these configurations, context graphs in the KG, text, and heterogeneous settings ($XG^K$, $XG^D$, and $XG^H$ respectively), are constructed in the manner described in Sec. 3.3. Uniqorn's XGs in the three setups are characterized in Table 5.

### 5.2.2. Baselines for the heterogeneous setup

We use UniK-QA [89], PullNet [111], and GRAFT-Net [112], as baselines for the KG+text heterogeneous setup. UniK-QA is the first system to explore answering over heterogeneous sources by verbalizing all inputs to text sequences, and is state-of-the-art in the prominent retrieve-and-read paradigm for QA over heterogeneous sources. More details below:

(i) **UniK-QA [89]**. UniK-QA verbalizes all KG facts to sentence form using simple rules, and adds them to the text corpus. This is then queried through a neural retriever (dense passage retriever, DPR [74]), and the top-100 evidences are fed into a generative reader model (fusion-in-decoder, FiD [64]). DPR was designed to work over Wikipedia: so we fine-tuned DPR for LC-QuAD 2.0 by using the subset of Wikipedia pages retrieved among the Google results containing a gold answer as positive instances, and randomly choosing a negative instance from the positive instance of a different question (see the original DPR paper [74] for details). We also fine-tune FiD, originally pre-trained using T5, on LC-QuAD 2.0. Some parts of the UniK-QA code is available publicly (DPR, FiD): the rest was reimplemented.

(ii) **PullNet [111]**. PullNet uses an iterative process to construct a question-specific subgraph for complex questions, where in each step, a relational graph convolutional neural network (R-GCN) is used to find subgraph nodes that should be expanded (to support multihop conditions) using "pull" operations on the KG and corpus. After the expansion, another R-GCN is used to predict the answer from the expanded subgraph. PullNet code is not public: it was completely reimplemented by the authors.

(iii) **GRAFT-Net [112]**. GRAFT-Net (Graphs of Relations Among Facts and Text Networks) also uses R-GCNs specifically designed to operate over heterogeneous graphs of KG facts and entity-linked text sentences. GRAFT-Net uses LSTM-updates for text nodes and directed score propagation via Personalized PageRank (PPR). GRAFT-Net code is available for the deprecated KG Freebase: non-trivial adaptations had to be made for answering over Wikidata. PullNet and GRAFT-Net need entity-linked text for learning their models, for which we used the same text corpora described earlier (10 documents per question from Google Web search) that was tagged with Tagme and Elq with the same $\rho$ threshold of zero (see Sec. 5.2.1).

### 5.2.3. KG-QA baselines

UniK-QA, PullNet, and GRAFT-Net can be run in KG-only modes as well. So these naturally add to KG-QA baselines for us. In addition, we use the systems QAnswer [54] and Platypus [117] as baselines for KG-QA. To the best of our knowledge, these are the only systems for Wikidata with sustained online services and APIs, with QAnswer having state-of-the-art performance. We use public APIs for both systems. QAnswer is a commercial and closed-source system that can also work over text of Wikimedia websites, but it cannot be made to run on corpora of our choice. Due to code unavailability, we could not evaluate text-only or heterogeneous variants of QAnswer. More details below:

(i) **QAnswer [36]**. This is an extremely efficient method that relies on an over-generation of SPARQL queries based on simple templates, that are subsequently ranked, and the best query is executed to fetch the answer. The queries are generated by a fast graph BFS (breadth-first search) algorithm relying on HDT indexing [43].

(ii) **Platypus [117]**. This was designed as a QA system driven by natural language understanding, targeting *complex questions* using grammar rules and template-based techniques. Questions are translated not exactly to SPARQL, but to a custom logical representation inspired by dependency-based compositional semantics.

[13] https://spacy.io/

[14] https://huggingface.co/bert-base-cased

[15] https://huggingface.co/docs/transformers/model_doc/bert#tfbertforsequenceclassification

#### 5.2.4. Text-QA baselines

The mainstream manifestation of Text-QA in the literature today is in the form of machine reading comprehension (MRC), where given a question and a corpus, the system derives an answer. There are many MRC systems today, along with open-retrieval variants where the question-relevant passages are not given but need to be retrieved from a large repository. As for Text-QA baselines in this work against which we compare Uniqorn, we focus on distantly supervised methods. These include neural QA models which are pre-trained on large question-answer collections, with additional input from word embeddings. These methods are well-trained for QA in general, but not biased towards specific benchmark collections. We are interested in robust behavior for ad hoc questions, to reflect the rapid evolution and unpredictability of topics in questions on the open Web. As specific choices, we adopt the well-known PathRetriever [6], DocumentQA [26], and DrQA [19] systems as representatives of robust open-source implementations, that can work seamlessly on unseen questions and passage collections. All methods can deal with multi-paragraph documents and multi-document corpora. They are deep learning-based systems with large-scale training via Wikipedia coupled with Text-QA benchmarks like NaturalQuestions [76], HotpotQA [136], SQuAD [97], and TriviaQA [70]: we use the pre-trained QA models on the benchmarks that these methods obtained their respective best performances on, and apply these models on our test sets. More details below:

(i) **PathRetriever [6]**. This method is specifically developed for *multi-hop complex questions*. The approach used is supervised iterative graph traversal (akin to PullNet [111]) and uses a novel recurrent neural network (RNN) method to learn to sequentially retrieve relevant passages in reasoning paths for complex questions, by conditioning on the previously retrieved documents. It is a very robust method that makes effective use of data augmentation and smart negative sampling to boost its learning, and has a minimal set of hyperparameters. The standard BERT QA model is used as the reader. Code was available from here[16]. The default configuration was used.

(ii) **DocumentQA [26]**. The DocumentQA system adapted passage-level reading comprehension to the multi-document setting. It samples multiple paragraphs from documents during training, and uses a shared normalization training objective that encourages the model to produce globally correct output. The DocumentQA model uses a sophisticated neural pipeline consisting of multiple CNN and bidirectional GRU layers, coupled with self-attention. For DocumentQA, we had a number of pre-trained models to choose from, and we use the one trained on TriviaQA (`TriviaQA-web-shared-norm`) as it produced the best results on our dev set. Defaults were used for other parameters[17].

(iii) **DrQA [19]**. This system combines a search component based on bigram hashing and TF-IDF matching with a multi-layer RNN model trained to detect answers in paragraphs. Since we do not have passages manually annotated with answer spans, we run the DrQA model pre-trained on SQuAD [97] on our test questions with passages from the ten documents as the source for answer extraction. Default configuration settings were used[18].

#### 5.2.5. Graph baselines

We compare our GST-based method to simpler graph algorithms based on breadth-first search (BFS) and ShortestPaths as answering mechanisms. More details below:

(i) **BFS [75]**. In BFS for graph-based QA, iterators start from each anchor node in a round-robin manner, and whenever at least one iterator from each anchor group meets at some node, it is marked as a candidate answer. At the end of 1000 iterations, answers are ranked by the number of iterators that met at the corresponding nodes.

(ii) **ShortestPaths**. As another intuitive graph baseline, shortest paths are computed between every pair of anchors, and answers are ranked by the number of shortest paths they lie on.

For fairness, BFS and ShortestPaths baselines are run on the same XGs and anchor nodes with which Uniqorn is applied. We perform additional fine-tuning for both BFS and ShortestPaths by finding the best thresholds for alignment insertion ($\tau^{E}_{align}$ and $\tau^{P}_{align}$) for these methods, using the development set.

### 5.3. Metrics

Most systems considered return exactly one answer, and so we use Precision@1 (P@1) as our metric. P@1 is a standard metric for factoid QA [99]: in several practical use cases like voice-based personal assistants, only one best answer can be returned to the user. The gold answer sets for all our benchmarks are entities and literals grounded (canonicalized) via the Wikidata KG. Thus, for KG-QA, evaluation is standard: answers are already KG entities and literals. When the top-1 answer matches any of the ground truths exactly, P@1 is 1, else 0. This is trickier for text, or KG+Text (when the answer source is text) settings, where the system response can be an arbitrary span of text, potentially containing abbreviations and noun phrases (*"The Rhine river"*, *"lotr"*, *"two million USD"*, *"St. Michael"*, etc.). This also applies for UniK-QA, that uses a generative model for formulating a textual answer, that may not be exactly found in any single source evidence. We tackle such cases as follows: when the returned answer span matches a Wikidata entity/literal in the ground truth set exactly, P@1 = 1. If no matches are found, we search for the string in KG entity aliases of the ground truth answers. If an exact

[16] https://github.com/AkariAsai/learning_to_retrieve_reasoning_paths

[17] https://github.com/allenai/document-qa

[18] https://github.com/facebookresearch/DrQA

**Table 6**
Basic statistics for the AMT study.

| | |
|---|---|
| Title | Verify correctness of machine-generated answers to fact-based questions |
| Description | Match machine-generated answers to a fact-based question, and the set of correct answers for the question. If you consider the pair to match, mark "yes", otherwise "no". When you are unsure, please consult Web Search. Cases to look out for are abbreviations, partial matches, and alternative formulations. |
| Participants | 91 unique Master Turkers |
| Time allotted per HIT | 120 minutes maximum |
| Time taken per HIT | 32 minutes on average |
| Annotations per QA pair | 1 |
| Questions in one HIT | 100 QA pairs (about 10 questions) |
| Payment per HIT | 7 USD |
| Total no. of questions | 6,923 (over all benchmarks, c.f. Table 3) |
| Total no. of QA pairs | 85,985 (over all benchmarks, systems, and setups) |
| Central tendency of #⟨Q, A⟩ pairs | 12.42 (mean), 10 (median), 10 (mode) |

match is found, then P@1 = 1, else we proceed to a *human evaluation* as discussed below.

### 5.4. Human evaluation

For the remaining cases where a system-generated answer did not exactly match a gold KG entity label or any of its aliases, we perform a crowdsourced human evaluation via Amazon Mechanical Turk (AMT). The task is straightforward: given a factoid question and an answer string, verify its correctness by matching against the gold answer set and optionally searching the Web in unsure cases. Textual response strings from all systems were pooled for every question to reduce annotation overhead. All string responses to a question like *Which river flows through Austria and Germany?* were grouped within an AMT HIT (Human Intelligence Task, a unit job on the platform). Each HIT consisted of 100 QA pairs. Responses from the different systems were shuffled with respect to their ordering in the AMT interface to avoid *position bias* in annotations. To prevent spam, only Master Turkers[19] from the US were allowed to participate. Workers were paid 15 USD per hour, which is fair compensation given the platform, consistent with the German minimum wages (all authors are affiliated to a German organization). One hour translated to slightly less than two HITs, where each HIT fetched 7 USD, and took slightly less than half an hour to complete. For all questions where a system-generated answer is deemed as correct by a Master worker, we assign P@1 = 1. For all remaining cases, we set P@1 = 0. On examining anecdotal cases, we found this human evaluation to be particularly worthwhile. The task is non-trivial – matching abbreviations (St. with Street or Saint), noun phrases (Rhine with The Rhine River), uncommon aliases ("Vol-kahn-o" with Oliver Kahn), and varying date formats, are some of the particularly tricky cases that would have made a completely automated machine evaluation a rather noisy alternative, leading to unreliable trends. Task details are in Table 6.

**Disclaimer**. Human evaluation is performed only on the **test sets** of all benchmarks, where it cost about 8000 USD for annotating approximately 86k QA pairs. On the dev set, one obtains several distinct answers for every question, as many different configurations are tested arising from hyperparameter variations. This makes human evaluation infeasible, or rather cost-exorbitant, as millions of unique QA pairs need to be assessed. For the dev set, we thus reward exact matches with any of the gold answers or their KG aliases, and partial matches when the matched answer word is a non-stopword. In these cases, P@1 = 1, otherwise 0. Also, this is another reason why other metrics popular in QA like MRR or Hit@5 could not be considered in this work as this would multiply the cost manifold even if only the top-5 answers for the test set questions over all systems, setups, and benchmarks had to be evaluated by humans.

## 6. Results and insights

In this section, we present our experimental results and the inferences drawn from these observations. We structure this section into key findings, in-depth analysis, and comparison with LLMs.

### 6.1. Key findings

Our main results are presented in Table 7. All tests of statistical significance hereafter correspond to the McNemar's test for paired binomial data, since P@1 is a binary metric. The significance level for all tests was set to 0.05 ($p$-value $< 0.05$ to be considered significant). Key findings below:

**(i) Uniqorn outperforms heterogeneous QA baselines**. Owing to its unique ability of harnessing multi-evidence knowledge in the collection via a combination of graph representation and graph algorithm, Uniqorn significantly beats baselines in heterogeneous QA (UniK-QA, PullNet, GRAFT-Net). The UniK-QA approach, that relies on a uniform verbalized representation of the evidences, is considered the state-of-the-art in heterogeneous QA. But it fails on the LC-QuAD 2.0 benchmark as it has no inherent support for dealing with more complex questions with multiple concepts and conditions. A key building block in UniK-QA, the fusion-in-decoder (FiD) generative reader that produces the final answers, can aggregate evidences over multiple sources in the likely sense of giving higher importance to evidences observed more than once (see usage in [24]). But it cannot join simpler evidences to reason for a more complex information need, which is one of our key contributions.

[19] https://www.mturk.com/help

**Table 7**
Comparison of Uniqorn and baselines over the LC-QuAD 2.0 test set, as measured by P@1 performance. The best value per column is in **bold**. An asterisk (*) indicates statistical significance of Uniqorn over the best baseline in that column. A hyphen ('-') indicates that the corresponding baseline cannot be applied to that particular setting.

| Method | KG+Text | KG | Text |
|---|---|---|---|
| Uniqorn (Proposed) | **0.324*** | **0.319** | 0.105 |
| UniK-QA [89] | 0.141 | 0.050 | 0.137 |
| PullNet [111] | 0.015 | 0.014 | - |
| GRAFT-Net [112] | 0.213 | 0.152 | - |
| QAnswer [36] | - | 0.315 | - |
| Platypus [117] | - | 0.042 | - |
| PathRetriever [6] | - | - | **0.212** |
| DocumentQA [26] | - | - | 0.175 |
| DrQA [19] | - | - | 0.098 |
| BFS [75] | 0.026 | 0.185 | 0.049 |
| ShortestPaths | 0.131 | 0.136 | 0.083 |

PullNet was one of the early systems for heterogeneous and complex QA. But it was developed exclusively for *multi-hop* complex questions (referred to as chain joins earlier in this text, or bridge questions [136]), and falls significantly short here as the system cannot handle questions that do not conform to a uniform multi-hop nature (varying number of hops, varying distribution of join types, or information being present within qualifiers). Such diversity of complex phenomena is a more realistic scenario and is indeed reflected in LC-QuAD 2.0 (unlike the MetaQA benchmark [140], for example), and in the smaller benchmarks considered later. Surprisingly, GRAFT-Net, a predecessor of PullNet, and sharing the same R-GCN node classifier as PullNet, turns out to be the strongest baseline in the heterogeneous mode. GRAFT-Net cannot handle arbitrary complex questions, but we attribute its success to being able to harness qualifier information (which is considered being in 1-hop of the question entity in GRAFT-Net). Here the question is no longer "simple", in the sense of being answerable via a single triple, but it is not multi-hop in the more conventional sense of the term either (*"In which Moscow stadium was the FIFA World Cup 2018 final played?"* is a complex question not answerable by a single Wikidata triple but a more complex KG fact with a qualifier).

Finally, the benchmark LC-QuAD 2.0 is a challenging one: it contains several complex questions that are beyond reach of all systems under test. With maximum performance of about 32%, it leaves plenty of room for improvement. Notably, the performance of Uniqorn is higher in the KG+Text case than for individual sources, a trend that is also seen in the three baselines discussed above – this can be attributed to a better information coverage provided by the heterogeneous case, a realistic setup for most Web search systems. This is thus a call to the community to develop more flexible systems not tailored to specific input sources.

For Uniqorn, there are 380 questions which could be answered by the KG+Text setup but not in the KG-only or Text-only setups, and in total Uniqorn (KG+Text) could answer $1,597$ questions out of the $4,921$ test cases. So in 24% cases (380/1597), we were able to exploit the mixture of sources when the individual sources both failed. For instance, *What is the name of the sister city tied to Kansas City, which is located in the county of Seville Province?* is one such question for which KG+Text setup succeeds in obtaining the correct answer whereas the KG-only or Text-only setups fail. This mainly happens because although the details of *"Kansas City"* and *"Seville Province"* are present in the KG, the notion of *"sister city"* is only found in the text corpus. From another perspective, if we look at all the *distinct* questions which are answered by Uniqorn for KG+Text, or KG, or Text, the number is $2,200$. Note that $2,200$ out of $4,921$ is about 44.7%, a reasonably high number for a difficult benchmark, and indicates that an appropriate ensemble of Uniqorn's answer derivations could lead to substantial gains over baselines. For UniK-QA, this number (distinct questions answered over all sources) is just 921 in contrast to Uniqorn's tally of $2,200$.

**(ii) Uniqorn maintains satisfactory performance on individual sources**. Uniqorn beats all baselines in KG-QA as well, and maintains a respectable performance on Text-QA. UniK-QA, the best competitor in terms of unified answering, does much worse on KG-QA but better for text sources. This can be explained by its Wikipedia-based training (also for its reader model FiD), that is more amenable to, and is inspired by, Text-QA. Text sources often have a complex information need all expressed within a single sentence (consider opening sentences of most Wikipedia articles), thus bringing several co-called complex questions in scope: this helps UniK-QA and other Text-QA methods. This is relatively less in KGs: while a qualifier makes a main fact richer with context information, such *n*-ary facts are rarely at the level of information density compared to Wikipedia opening sentences. Compound KG facts are also not that easy to exploit, requiring sophisticated querying (compared to highly advanced methods for neural matching of a question and a complex NL sentence). GRAFT-Net and PullNet, the other heterogeneous QA baselines, unfortunately cannot work in pure text-only modes: they need entity linking and shortest KG paths during training, and can only return crisp entities as answers, for which embeddings have been learnt. Uniqorn does not suffer from these drawbacks. The superior performance of MRC models (PathRetriever, DocumentQA, DrQA) in Text-QA can be attributed to powerful neural models. Uniqorn has a much simpler pipeline that is rather designed for seamless QA over several input sources. Moreover, the strict graph/structure induction from raw text may not be ideal in every scenario. The XG construction phase is rather noisy, where candidate answers may not be captured by the open IE extractor. MRC methods lack a need for entity-based representations and can return arbitrary spans of text from the context passages as answers: this substantially increases their performance on Text-QA benchmarks.

**Table 8**
Representative examples from the LC-QuAD 2.0 test set where UNIQORN was able to compute a correct answer at the top rank (P@1 = 1), but none of the baselines could.

| KG+Text | KG | Text |
|---|---|---|
| *Who is the parent agency of the maker of Novo Nordisk (United States)?* | *Who is the husband of the child of Emmanuel Bourdieu?* | *Tell me which collectible card game played with a computer keyboard has the highest number of players?* |
| *What prize money did Wangari Maathai receive for the Nobel Peace Prize?* | *Give me the end time for Martina Navratilova as a member of sports team as Czechoslovakia Federation Cup team?* | *What is the cation with the highest isospin quantum number?* |
| *Which work earned Ron Howard the Primetime Emmy Award for Outstanding Miniseries?* | *What is the street number of Musee D'Orsay has located on street as Rue De Bellechasse?* | *What disease has the shortest incubation period in humans?* |
| *Name the boyfriend of Aphrodite who has the child of Cephalus.* | *What character did actor Richard Williams perform the voice of in Who Framed Roger Rabbit?* | *Who are the fictional characters in Nineteen Eight-Four?* |
| *What method of murder is mentioned in the dedication of Church of St Peter and St Paul, in the Church of England?* | *Name the mother and the date on which she gave birth to Princess Maria Amélia of Brazil?* | *Who received the Academy Award for Best Art Direction, Black and White for directing Sunset Boulevard?* |

**(ii) Computing GSTs on a question-specific context graph is a preferable alternative to SPARQL-based systems over RDF graphs**. This is seen with the superior performance of UNIQORN over QANSWER and PLATYPUS, the two SPARQL-based systems. Let us examine this in more detail. For KGs, QANSWER is clearly the stronger baseline. It operates by mapping names and phrases in the question to KG concepts, enumerating all possible SPARQL queries connecting these concepts that return a non-empty answer, followed by a query ranking phase. This approach works really well for relatively simpler questions with a few keywords. However, the query ranking phase becomes a bottleneck for complex questions with multiple entities and relations, resulting in too many possibilities. This is the reason why reliance on the best SPARQL query may be a bad choice. Additionally, unlike UNIQORN, QAnswer cannot leverage any qualifier information in Wikidata. This is both due to the complexity of the SPARQL query necessary to tap into such records, as well as an explosion in the number of query candidates if qualifier attributes are allowed into the picture. Our GST establishes a common question context by joint disambiguation of all question phrases, and smart answer ranking helps cope with the noise in the XG later. While QAnswer is completely syntax-agnostic, Platypus is the other extreme: it relies on accurate dependency parses of the NL question and hand-crafted question templates to come up with the best logical form. This performs impressively when questions fit the supported set of syntactic patterns, but is brittle when exposed to a multitude of open formulations from an ensemble of QA benchmarks. The relatively better performance of GRAFT-NET in KG-QA also attests to the superiority of graph-based search over SPARQL for complex questions.

**(iii) Graph-based methods open up the avenue for this unified answering process**. Searching for optimal interconnections between anchors via GSTs is essential and powerful. Use of GSTs is indeed required for best performance, and cannot be easily approximated with simpler graph methods like BFS and SHORTESTPATHS. It is notable that these relatively simple graph baselines maintain a consistent performance across benchmarks and sources (Tables 7 and 9), compared to more sophisticated methods. This can be explained by the fact that paths between entity pairs often create a compact zone in the graph, that acts like a faster and noisier approximation for GSTs.

**Takeaway**. Holistically, all our main findings point to the high potential of graph-based methods for tackling the challenging problem of answering complex questions over heterogeneous sources. We show representative examples from the various benchmarks in Table 8 where only UNIQORN was able to return the correct answer, to give readers a feel of the complexity of information needs that is in our scope.

## 6.2. In-depth analysis

We now present a wide variety of additional results that compare and contrast each and every aspect of the UNIQORN's pipeline in detail.

**Zero-shot QA experiments**. Most entity-oriented QA systems today need trained entity and predicate embeddings at inference time. As a result, they cannot deal with the case when the test set contains entities or predicates unseen during training. This is a major disadvantage that makes such systems rather benchmark-specific, and most benchmarks often contain entities and predicates in test questions that were parts of some other questions in the train set. Hence,

**Table 9**

**Zero-shot QA experiments:** Comparison of P@1 performance of UNIQORN and baselines on the smaller LC-QuAD 1.0, ComQA, CQ-W, CQ-T, and QALD datasets, where the models trained on LC-QuAD 2.0 were directly run on the questions in these five benchmarks. The best value per column is in **bold**. An asterisk (*) indicates statistical significance of UNIQORN over the best baseline in that column. '-' indicates that the corresponding baseline cannot be applied to this setting.

| Methods (KG+Text) | LCQ1 | ComQA | CQ-W | CQ-T | QALD |
|---|---|---|---|---|---|
| UNIQORN (Proposed) | **0.265*** | **0.243** | **0.320*** | **0.393*** | **0.157** |
| UNIK-QA [89] | 0.132 | 0.114 | 0.040 | 0.040 | 0.143 |
| PULLNET [111] | 0.019 | 0.010 | 0.000 | 0.013 | 0.000 |
| GRAFT-NET [112] | 0.084 | 0.030 | 0.047 | 0.007 | 0.000 |
| QANSWER [54] | - | - | - | - | - |
| PLATYPUS [117] | - | - | - | - | - |
| PATHRETRIEVER [6] | - | - | - | - | - |
| DOCUMENTQA [26] | | - | - | - | - |
| DRQA [19] | - | - | - | - | - |
| BFS [75] | 0.035 | 0.005 | 0.020 | 0.007 | 0.014 |
| SHORTESTPATHS | 0.162 | 0.228 | 0.187 | 0.253 | 0.129 |
| **Methods (KG)** | **LCQ1** | **ComQA** | **CQ-W** | **CQ-T** | **QALD** |
| UNIQORN (Proposed) | **0.170** | 0.099 | 0.187 | 0.107 | **0.129** |
| UNIK-QA [89] | 0.042 | 0.045 | 0.013 | 0.000 | 0.029 |
| PULLNET [111] | 0.016 | 0.040 | 0.000 | 0.000 | 0.000 |
| GRAFT-NET [112] | 0.053 | 0.045 | 0.033 | 0.007 | 0.000 |
| QANSWER [54] | 0.164 | **0.238** | **0.287** | **0.147** | 0.114 |
| PLATYPUS [117] | 0.021 | 0.010 | 0.007 | 0.007 | 0.100 |
| PATHRETRIEVER [6] | - | - | - | - | - |
| DOCUMENTQA [26] | - | - | - | - | - |
| DRQA [19] | - | - | - | - | - |
| BFS [75] | 0.117 | 0.054 | 0.107 | 0.080 | 0.014 |
| SHORTESTPATHS | 0.077 | 0.074 | 0.133 | 0.047 | 0.057 |
| **Methods (Text)** | **LCQ1** | **ComQA** | **CQ-W** | **CQ-T** | **QALD** |
| UNIQORN (Proposed) | 0.214 | 0.183 | 0.213 | 0.307 | 0.243 |
| UNIK-QA [89] | 0.136 | 0.188 | 0.113 | 0.053 | 0.100 |
| PULLNET [111] | - | - | - | - | - |
| GRAFT-NET [112] | - | - | - | - | - |
| QANSWER [54] | - | - | - | - | - |
| PLATYPUS [117] | - | - | - | - | - |
| PATHRETRIEVER [6] | **0.327** | 0.564 | 0.527 | 0.320 | **0.271** |
| DOCUMENTQA [26] | 0.290 | **0.599** | **0.653** | **0.520** | 0.243 |
| DRQA [19] | 0.141 | 0.168 | 0.280 | 0.400 | 0.029 |
| BFS [75] | 0.077 | 0.114 | 0.147 | 0.167 | 0.086 |
| SHORTESTPATHS | 0.130 | 0.208 | 0.167 | 0.233 | 0.114 |

this effect of unseen items at answering time is often not noticeable anymore. However, zero-shot QA needs to be as free as possible from the constraints of benchmarks. As a result, we designed a novel zero-shot QA experiment where models trained on our larger and main benchmark LC-QuAD 2.0 are directly applied off-the-shelf to the other five, smaller benchmarks (LC-QuAD 1.0, ComQA, CQ-W, CQ-T, and QALD) without any further training or parameter tuning. The benchmarks LC-QuAD 2.0 and LC-QuAD 1.0 were created independently, so there is no overlap between the LC-QuAD 2.0 train set and LC-QuAD 1.0 test set (or any of these five test sets).

**Table 10**

Pipeline ablation results on the LC-QuAD 2.0 dev set with P@1. Best values per column are in **bold**. Statistically significant differences from the full configuration are marked with *.

| Configuration | KG+Text | KG | Text |
|---|---|---|---|
| Full | **0.292** | 0.266 | 0.108 |
| No entity alignment | 0.289 | - | 0.101 |
| No predicate alignment | 0.240* | - | 0.097 |
| No type alignment | 0.286 | - | 0.110 |
| No type nodes | 0.289 | **0.279** | **0.114** |

Results are presented in Table 9. The main observation is that UNIQORN does not need known embeddings of entities and predicates, and hence **outperforms all baselines across all benchmarks in the heterogeneous setup** without any additional training effort. Notably, GRAFT-NET and PULLNET, otherwise very strong systems, suffer from this limitation of unseen entities and predicates at inference time, and hence display substantially worse results in these zero-shot experiments as compared to the previous closed setup with LC-QuAD 2.0 with dedicated train, dev and test splits (Table 7). UNIK-QA and the Text-QA baselines adopting the retriever-reader architecture do not have this drawback, and perform better than GRAFT-NET and PULLNET. This effect is seen in KG+Text and KG setups: Text-QA trends remain mostly the same as before, where PATHRETRIEVER and DOCUMENTQA outperform UNIQORN. Notably, UNIK-QA performs systematically worse than UNIQORN in the text-only mode on most benchmarks, except for ComQA. A possible reason could be that the questions in ComQA are originally derived from WikiAnswers (a deprecated Community QA platform), a pure text collection. The UNIK-QA pipeline, owing to its reliance on text-oriented verbalization, works better in such a scenario. It is worthwhile to note that simpler unsupervised systems often shine in zero-shot QA: QANSWER wins in the KG setup on three out of five datasets, while the naive SHORTESTPATHS baseline achieves a consistently respectable performance for KG+Text QA.

**Ablation experiments**. To get a proper understanding of UNIQORN's robustness, it is important to systematically ablate its configurational pipeline (Table 10). We do not insert alignment edges in the KG-only mode as KG items are already canonicalized: these are marked by the hyphens in the corresponding slots. We note the following:

- Removing entity alignment consistently degrades performance (compare Row 1 with Row 2).
- Removing predicate alignment edges also systematically reduces P@1, and even more strongly than that for entities (compare Row 1 with Row 3, significant drop in KG+Text column).
- We noted a drop in P@1 from 0.292 to 0.286 for type alignment edges (Row 1 vs. Row 4) and 0.292 to 0.289 for type nodes (Row 1 vs. Row 5) in the heterogeneous setup, our primary focus. Thus, while it appears that types did not help for the individual sources, to keep our configuration uniform, we decided to keep type nodes and alignments in all the three scenarios.

**Open IE alternatives**. UNIQORN has a number of text and KG preprocessing steps that warrant a more in-depth look. In particular, the Open IE for UNIQORN is tailored to our unique needs of extracting as many crisp triples as possible from text, while allowing for a certain amount of noise. Previously we experimented with Stanford's Open Information Extraction library [5] as an alternative to the customized triple extractor of UNIQORN. However, the results were very poor as Stanford's conservative extractor is geared towards precision rather than recall, missing out on many relevant pieces of evidence. We had also tried ClausIE [32] and OpenIE 5.0 tools [87], but both produced very long object phrases in order to cover most words in a sentence. Such verbose outputs are too noisy and unsuitable for downstream processing, as we need to stitch shared subjects and objects together to construct the graph.

Finally, we explored the recently released alternative Text2AMR2FRED [48, 49] based on AMR [8], which produces triples containing both surface forms of question tokens, and those connecting the question to its abstract meaning representation (AMR). We replaced triples obtained by our own extractor with those obtained from Text2AMR2FRED (in NTriples format, accessed via the API[20]) on the same top-$k$ evidence pieces, and plugged them into the UNIQORN pipeline. The RDF URI/namespace prefixes were stripped from the triples to turn them into crisp text. This alternative resulted in slightly better QA performance (P@1 of 0.111 on our dev set of 1000 questions, compared to UNIQORN's 0.108, for Text-QA), but this difference was not statistically significant.

On the downside, Text2AMR2FRED has up to 10$x$ longer run-times (UNIQORN averages 0.308 seconds per question for the corresponding pipeline steps, compared to 3.7 seconds for Text2AMR2FRED). The reason for this efficiency drawback is that it yields a much larger set of triples as it also aims to cover in-depth semantics of sentences (the average number of triples generated from each snippet is 16.01 for UNIQORN but 174.70 for Text2AMR2FRED). These are useful for other use cases, but our QA setup is already served well by a smaller set of informative triples with surface tokens. Overall, alternative tools like Text2AMR2FRED are an interesting route to explore for future QA, but does not offer many advantages over our customized IE techniques.

**Text-processing components**. To obtain insights on the robustness of our text-processing components, we analyzed the performance of the coreference resolution (CR), named entity recognition (NER), and named entity disambiguation (NED) employed by UNIQORN. We sampled a set of 50 sentences (for 50 different questions) from the Text-QA pipeline of UNIQORN, such that the sentences contain instances of the noisy CR (our rule of replacing a pronoun with nearest preceding entity, along with some adjustments) and NER with spaCy. NED is not performed for Text-QA, so we did this analysis for KG-QA instead. We sampled 50 cases in total, 25 cases for ELQ and 25 for TAGME, the two NED systems in UNIQORN. We manually evaluated these for correctness, and obtained the following results: (i) 70% cases of CR were correct, and (ii) 90% cases of NER were correct. For entity linking to the KG, 64% of the cases were correct: 80% for ELQ and 48% for TAGME. While having high recall, TAGME has lower NED accuracy but it also retrieves many more entities than ELQ (for 25 sample questions, TAGME links 140 entities, compared to 72 for ELQ). Since UNIQORN takes the union of the two systems' outputs and employs various noise-mitigation techniques, TAGME's lower accuracy does not pose any problems. This indicates that the intermediate processing steps of UNIQORN have decent accuracy. Still, our design choices primarily aim for efficiency, trading off a small loss in quality.

[20] https://arco.istc.cnr.it/txt-amr-fred/api/docs

**Table 11**
BERT fine-tuning results on LC-QuAD 2.0 dev set with P@1.

| Training time | 1 | 3 | 5 | 10 | 20 |
|---|---|---|---|---|---|
| KG+Text | 7h 46m | 16h 47m | 22h 53m | 45h 40m | 78h 42m |
| KG | 3h 53m | 6h 57m | 9h 38m | 14h 3m | 18h 31m |
| Text | 3h 50m | 9h 5m | 11h 55m | 20h 55m | 39h 49m |
| **Classif. accuracy** | **1** | **3** | **5** | **10** | **20** |
| KG+Text | 0.919 | 0.944 | 0.959 | 0.977 | 0.986 |
| KG | 0.974 | 0.983 | 0.986 | 0.989 | 0.991 |
| Text | 0.849 | 0.899 | 0.931 | 0.957 | 0.974 |

**Table 12**
Different answer ranking results on the LC-QuAD 2.0 dev set on P@1. Best values per column are in **bold**. Statistically significant drops from the first row are marked with *.

| Answer ranking | KG+Text | KG | Text |
|---|---|---|---|
| GST counts | **0.292** | **0.266** | **0.108** |
| GST costs | 0.282 | **0.266** | 0.094 |
| GST node weights | 0.284 | 0.264 | 0.098 |
| Anchor distances | 0.225* | 0.250* | 0.082* |
| Weighted anchor distances | 0.227* | 0.260 | 0.082* |

**BERT fine-tuning**. We conducted an ablation study to quantify the effect of the number of negative instances per positive instance for fine-tuning the BERT evidence classifier. We varied the number of negative samples as 1, 3, 5, 10, 20 (roughly doubling with each step) per positive sample, and measured the training time and classification accuracy for each choice. The results are shown in Table 11. So we see that going beyond five negative samples massively increases the training time. For KG+Text, our primary setup, processing $10-20$ negative instances per positive sample takes almost two days or more. Since the classification accuracy achieved is already 93.1% to 98.6% with five negative samples, we believe that going beyond five negatives is not necessary. Note that our validation (dev) set for BERT fine-tuning contains a 1:1 ratio of positive:negative cases, so that this ballpark for accuracy really means that the classifier is working very well.

**Answer ranking**. UNIQORN ranks answers by the number of different GSTs that they occur in, the more the better (Row 1

**Table 13**
Effect of multi-document evidence shown via edge contributions by distinct documents to GSTs for the LC-QuAD 2.0 test set.

| | KG+Text | | KG | | Text | |
|---|---|---|---|---|---|---|
| GST Ranks | Avg. #Docs in GST | % of Questions with Answers in GST | Avg. #Docs in GST | % of Questions with Answers in GST | Avg. #Docs in GST | % of Questions with Answers in GST |
| 1 | 2.43 | 42.56% | N. A. | 39.20% | 1.43 | 15.89% |
| 2 | 2.45 | 42.92% | N. A. | 40.87% | 1.47 | 16.18% |
| 3 | 2.46 | 43.16% | N. A. | 41.78% | 1.50 | 16.36% |
| 4 | 2.49 | 43.45% | N. A. | 42.61% | 1.53 | 16.58% |
| 5 | 2.50 | 43.69% | N. A. | 43.40% | 1.55 | 16.85% |
| #Evidences in GST | Avg. Rank of GST | % of Questions with Answers in GST | Avg. Rank of GST | % of Questions with Answers in GST | Avg. Rank of GST | % of Questions with Answers in GST |
| 1 | 4.26 | 3.58% | N. A. | N. A. | 3.33 | 2.97% |
| 2 | 3.48 | 16.56% | N. A. | N. A. | 3.48 | 6.28% |
| 3 | 3.20 | 17.58% | N. A. | N. A. | 3.51 | 5.28% |
| 4 | 3.07 | 6.10% | N. A. | N. A. | 3.75 | 2.50% |
| 5 | 2.87 | 1.10% | N. A. | N. A. | 3.83 | 0.63% |

in Table 12). However, there arise some natural variants that would leverage a weighted sum rather than a simple count. This weight can come from the following sources: (i) the total cost of the GST, the lower the better [Row 2]; (ii) the sum of node weights in the GST, reflecting overall question relevance, the higher the better [Row 3]; (iii) total distances of answers to the anchors in the GST, the lower the better [Row 4]; and, (iv) a weighted version of anchor distances, when edge weights are available, the lower the better [Row 5]. We find that using plain GST counts (analogous to a simple voting strategy) works surprisingly well, leading to its use in the final UNIQORN pipeline. Zooming in to the GST and examining proximities of answers to anchors is not necessary. It even hurts significantly for most of the setups, marked by an asterisk (*).

**Joining multiple evidences**. For heterogeneous and KG-QA, UNIQORN outperformed all baselines, and in Text-QA, UNIQORN improved over DRQA on LC-QuAD 2.0, even though the latter is a supervised deep learning method trained on the large TriviaQA dataset. One of the reasons behind this success of UNIQORN is the ability of joining fragments of evidence across documents via GSTs. UNIQORN benefits two-fold from multi-document evidence:

- Confidence in an answer increases when all conditions for correctness are indeed satisfiable (and found) when looking at multiple documents. This increases the answer's likelihood of appearing in *some GST*.
- Confidence in an answer increases when it is spotted in multiple documents. This increases its likelihood of appearing in the *top-k* GSTs, as presence in multiple documents increases weights and lowers costs of the corresponding edges.

Details on the use of multi-document information is in Table 13. We make the following observations:

- Looking at the "Avg. #Docs in GST" columns in the upper half of the table, we see that considering the top-10 GSTs is worthwhile as all the bins combine evidence from multiple ($2-3$ on average) documents. This is measured by labeling edges in GSTs with documents (doc-ids) that contribute the corresponding edges.
- Moreover, they also contain the correct answer uniformly often (corresponding "#Questions with Answers in GST" columns). This is about 43%, 40%, and 16% for KG+Text, KG and Text respectively. The numbers in these columns do not add up to 100% because: (i) they are computed over the full test set and not the questions for which answers could be located in GSTs; and (ii) there are multiple correct answers for the same question, and they can be spread over differently ranked GSTs.
- The bottom half of the table inspects the inverse phenomenon, and finds that considering only the top-1 GST is not sufficient for aggregating multi-document evidence. GSTs combining multiple documents can be found beyond rank 1, and they often contain correct answers. For example, GSTs combining three documents are on average at rank 3 for KG+Text, and contain the answer 17.56% of the time (far from negligible).
- Finally, there is a sweet spot for GSTs aggregating nuggets from multiple documents to contain correct answers, and this turns out to be around three documents (see corresponding "#Questions with Answers in GST" columns). This, however, is an effect of the nature of questions in our benchmarks, that are not complex enough to require stitching evidence from more than three documents.

Most statistics for the KG setup are 'N. A.' (not applicable) because we do not treat different facts as coming from different document sources. Next, the heterogeneous setup necessitates a key measurement, that how many times, in fact, the top GSTs combine evidence from the two sources. This is reported in Table 14. A GST is said to mix information from KG and Text when it has either a node or an edge from each of the sources. About 80% top-1 GSTs and $10-16\%$ lower-ranked GSTs contain nodes or edges from both sources: this is concrete proof that UNIQORN indeed captures information from multiple sources for answering complex questions.

**Table 14**
Source heterogeneity in top-ranked GSTs in KG+Text setup.

| GST Rank | Both KG and Text Presence |
|---|---|
| 1 | 80.61% |
| 2 | 10.59% |
| 3 | 10.59% |
| 4 | 14.00% |
| 5 | 16.89% |

**Table 15**
Percentages of different error scenarios where the answer is not at the top-1 position, averaged over the LC-QuAD 2.0 test set.

| Error scenario | KG+Text | KG | Text |
|---|---|---|---|
| Ans. not in orig. evi. pool | 33.99% | 46.78% | 53.00% |
| Ans. in orig. evi. pool but not in XG | 30.78% | 27.43% | 27.76% |
| Ans. in XG but not in top-10 GSTs | 20.89% | 4.69% | 13.90% |
| Ans. in top-10 GSTs but not in top-5 cand. | 8.71% | 7.17% | 4.60% |
| Ans. in top-5 cand. but not at top-1 | 5.62% | 13.94% | 0.73% |

Ans.: Answer, orig.: original, evi.: evidence, cand.: candidates

**Error analysis**. In Table 15, we extract all questions for which UNIQORN produces an imperfect ranking (P@1 = 0), and discuss cases in a cascaded style. Each column adds up to 100%, and reports the distribution of errors for the specific setting. Note that values across rows are not comparable. Trends are comparable across input sources. We make the following observations:

(i) **[Row 1]** indicates the sub-optimal retrieval via Google (from the Web), and NED systems TAGME and ELQ (from Wikidata) with respect to complex questions. Strictly speaking, this is out-of-scope for UNIQORN. Nevertheless, an ensemble of search engines (Google + Bing) or NED systems (TAGME + ELQ + CLOCQ) may help improve answer coverage for UNIQORN.

(ii) **[Row 2]** indicates answer presence in $K^H$ or $K$ or $K^D$, but not in the respective $XG^H$ or $XG^K$ or $XG^D$. This is an effect of BERT-fine tuning errors made in the relevance assessments with respect to the question.

(iii) **[Row 3]** Presence of an answer in the XG but not in top-10 GSTs usually indicates an incorrect anchor matching, or sub-optimal GST ranking. This could be because one of the entities detected by the NED systems could be erroneous, or an irrelevant question phrase became an anchor. Anchor detection uses KG aliases, that are often incomplete. Revisiting the edge scoring mechanism in GSTs (instead of directly relying on BERT scores), or incorporating both node and edge weights into GST scoring, could also improve eventual GST ordering.

(iv) **[Rows 4 and 5]** represent cases when the answer is in the top-10 GSTs but languish at lower ranks in the candidates. Exploring weighted rank aggregation by tuning on the development set with variants in Table 12 is a likely remedy. A high volume of errors in this bucket is actually one of positive outlook: the core GST algorithm generally works well, and significant performance gains can be obtained by fine-tuning the ranking function with additional parameters.

**Implementation details**. All code is in Python, making use of the popular PyTorch library[21]. Whenever a neural model was used, code was run on a single NVIDIA Quadro RTX 8000 GPU with 48 GB GDDR6 memory. All code, data, and results for UNIQORN are publicly available at `https://uniqorn.mpi-inf.mpg.de`.

**Runtime analysis**. To conclude our detailed introspection into the inner workings of UNIQORN, we provide a distribution of runtimes of UNIQORN and all baselines in the three setups, over the 1000 dev set questions (Table 16). Training times of methods (where applicable) are not counted as training is assumed to be done offline. Fortunately, the use of the fixed-parameter tractable exact algorithm for GSTs helps achieve relatively short completion times for the core GST step ("Group Steiner Tree computation on XG" step from Fig. 2) for a large number of questions (about 5 seconds for KG+Text and sub-second for KG, Text: Row 7 in Table 16). Related graph algorithms like BFS and SHORTESTPATHS (last two rows), that could be viewed as approximations of GSTs, are accordingly a bit faster (about 3 seconds for KG+Text and sub-second for KG, Text) at the cost of reduced answering performance (cf. Table 7). However, end-to-end answering times of UNIQORN are still rather high for KG and KG+Text setups (Row 9). One caveat behind the fast runtimes for our main competitor UNIK-QA is that it assumes encodings of all evidences to be directly available at inference time, which is not the case in UNIQORN. Almost all the overhead for UNIQORN is due to KG processing (in the "triple extraction" and "on-the-fly XG construction" steps from Fig. 2): fact and type lookups, using KG shortest paths for injecting connectivity into the XGs, and scoring a large number of facts with BERT. BERT scoring takes about $51-65\%$ of the total runtimes in KG+Text QA or KG-QA (Row 4 in Table 16). This is negligible for Text-QA due to the relatively fewer triples to score. Inserting alignment edges also takes a substantial proportion of UNIQORN's total time (about 5 seconds for Text-QA, and 40 seconds for KG+Text), as it involves a large number of pairwise similarity computations. The final answer scoring step in all setups takes only a few milliseconds ("answer scoring" step from Fig. 2). Improving UNIQORN's total runtime is promising future work: ideas include parallelization of BERT encoding, using LLMs with fewer parameters like TinyBERT [68], and smart hashing algorithms for similarity computations.

**GSTs contribute to explainability**. Finally, we posit that Group Steiner Trees over question-focused context graphs help in understanding the process of answer derivation for an end-user. We illustrate this using three anecdotal examples of GSTs, one each for KG+Text-QA, KG-QA, and for Text-QA, in Figs. 5a through 5c. The corresponding question is in the respective caption. Anchor nodes are underlined and answers are in blue. The detailed legend is in the figure caption. An interesting thing to note for KG+Text is the co-existence of canonicalized and relaxed entities (`erythromycin`, *"many medicines"*) and predicates (`significant drug interaction`, *"metabolism of"*).

[21] `https://pytorch.org`

**Table 16**
Per-question answering times on average (in seconds) of methods over the LC-QuAD 2.0 dev set.

| Method | KG+Text | KG | Text |
|---|---|---|---|
| Uniqorn (NERD) | 4.316 | 4.316 | 0.000 |
| Uniqorn (KG lookups with Clocq) | 50.762 | 50.762 | 0.000 |
| Uniqorn (Triple extraction from text snippets) | 1.216 | 0.000 | 1.216 |
| Uniqorn (Scoring triples with BERT) | 105.552 | 104.893 | 0.659 |
| Uniqorn (Alignment computation) | 40.470 | 0.000 | 4.968 |
| Uniqorn (XG postprocessing) | 0.089 | 0.033 | 0.056 |
| Uniqorn (GST computation) | 4.569 | 0.017 | 0.248 |
| Uniqorn (Answer ranking) | 0.038 | 0.002 | 0.004 |
| Uniqorn (Full pipeline) | 207.010 | 160.022 | 7.151 |
| QAnswer [36] | - | 7.658 | - |
| Platypus [117] | - | 9.912 | - |
| PathRetriever [6] | - | - | 8.077 |
| DocumentQA [26] | - | - | 16.67 |
| DrQA [19] | - | - | 213.17 |
| UniK-QA [89] | 2.429 | 0.305 | 2.124 |
| PullNet [111] | 100.058 | 98.309 | - |
| GRAFT-Net [112] | 0.524 | 0.392 | - |
| BFS [75] | 301.939 | 258.843 | 7.567 |
| ShortestPaths | 305.093 | 258.844 | 7.821 |

## 6.3. Comparison with LLMs

**Setup**. Although large language models (LLMs) are not really comparable to Uniqorn in terms of functionality and data assets, this subsection discusses some experiments that shed light into the relative strengths and weaknesses of the LLM and graph-based architectures. Uniqorn proceeds in two steps of evidence retrieval and answer generation: it thus resembles RAG (Retrieval Augmented Generation) [79] enhancement for LLMs. So we compare the results of Uniqorn with LLMs for both zero-shot and RAG settings. To this end, we conducted experiments with OpenAI's GPT-4o[22] and Microsoft's Phi-3[23]. GPT-4o (released May 2024) is widely viewed as the world's most successful LLM overall, but it is proprietary and closed source, and can only be accessed via a paid API or through the ChatGPT interface. On the other hand, Microsoft's Phi-3 (updated June 2024) is one of the most recent and best-performing open instruction-tuned LLMs that can be run on an edge device like a laptop (we used the 3.8B parameters version, `Phi-3-mini-4k-instruct`). Due to cost and resource constraints, we ran these exploratory experiments on the 1000-question dev set. Interestingly, we found that Uniqorn significantly outperformed both GPT-4o and Phi-3 in a RAG setup.

**Configuration**. In the zero-shot setting, LLMs could solely rely on their parametric memory from pre-training and fine-tuning. In the RAG setting, LLMs were allowed to access to the top-$k$ BERT-retrieved evidence snippets of UNIQORN (via prompts), to generate an answer.

[22] `https://platform.openai.com/docs/models/gpt-4o`
[23] `https://huggingface.co/microsoft/Phi-3-mini-4k-instruct`

We used the following prompt in the **zero-shot setting**: *"Please provide a crisp answer to the following question. Your response should ideally be short strings (or lists of short strings). These strings could be entity labels of names and places, numbers, dates or other strings like quotations, etc.* `[question]`*."*

The following prompt was used in the **RAG setting** (the top-$k$ BERT evidences were appended to the prompt as *context*): *"Please provide a crisp answer to the following question. Your response should ideally be short strings (or lists of short strings). These strings could be entity labels of names and places, numbers, dates or other strings like quotations, etc.* `[question]`. *You can only use this specified context for answering:* `[context]`. *If the provided context does not contain the answer, please output the following string:* `The given evidence does not contain the answer to the input question`. *Please do not generate answers from your parametric memory or world knowledge."*

We accessed GPT-4o through its API. The maximum number of output tokens was set to 50 (ca. 200 characters) to limit responses to crisp entity names and similar strings (or short lists of these). A temperature of 0.0 was used for deterministic responses. A context limit of 25k characters was used to adhere to the API's rate limits (about 8k tokens); this is a reasonable choice for most prompts, including the RAG setup (the context length was exceeded in only about 30 out of 1000 cases for RAG over KG+Text, the setting with the longest prompts). A local installation of Phi-3 was used, with its default configuration.

**Evaluation**. To evaluate the answers from GPT-4o and Phi-3, we used a separate GPT-4o agent that compared the LLM-generated answers to the gold answers in the benchmark.

**Figure 5:** GSTs construct interpretable evidence for complex questions. Legend: Green circle = Entity node; Blue circle = Answer entity node; Diamond = Predicate node; Triangle = Type node; Line with label = Edge labeled by type = {Triple, Type, Alignment (dashed)}; Item with underline = Anchor node. (color online)

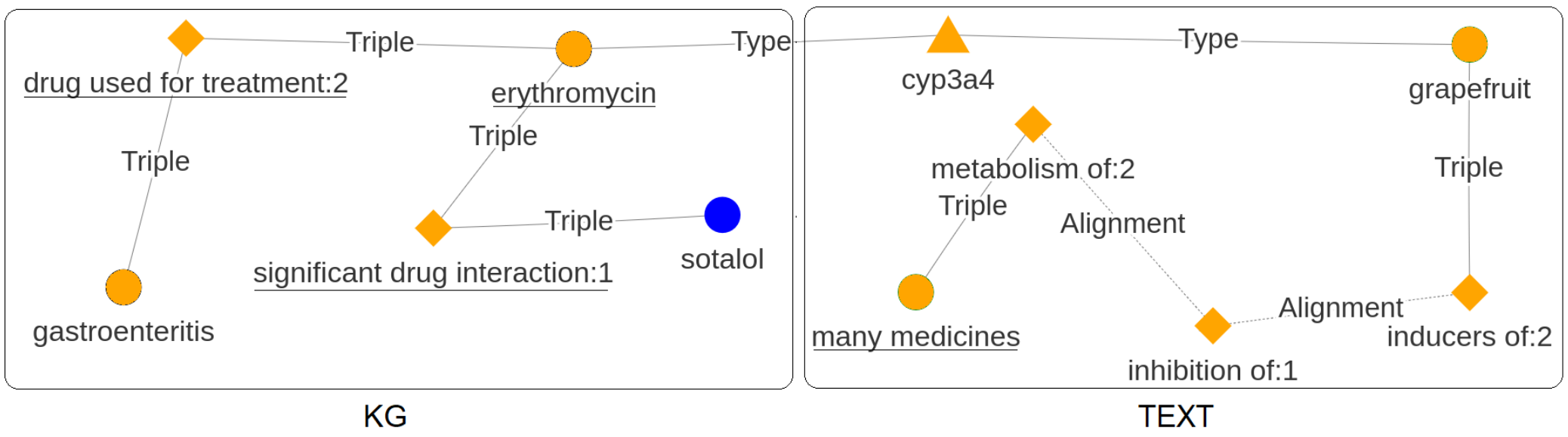

(a) Example GST for *Which medicines have a major drug interaction with erythromycin?* in KG+Text setup. The KG and Text parts are enclosed in illustratory boxes.

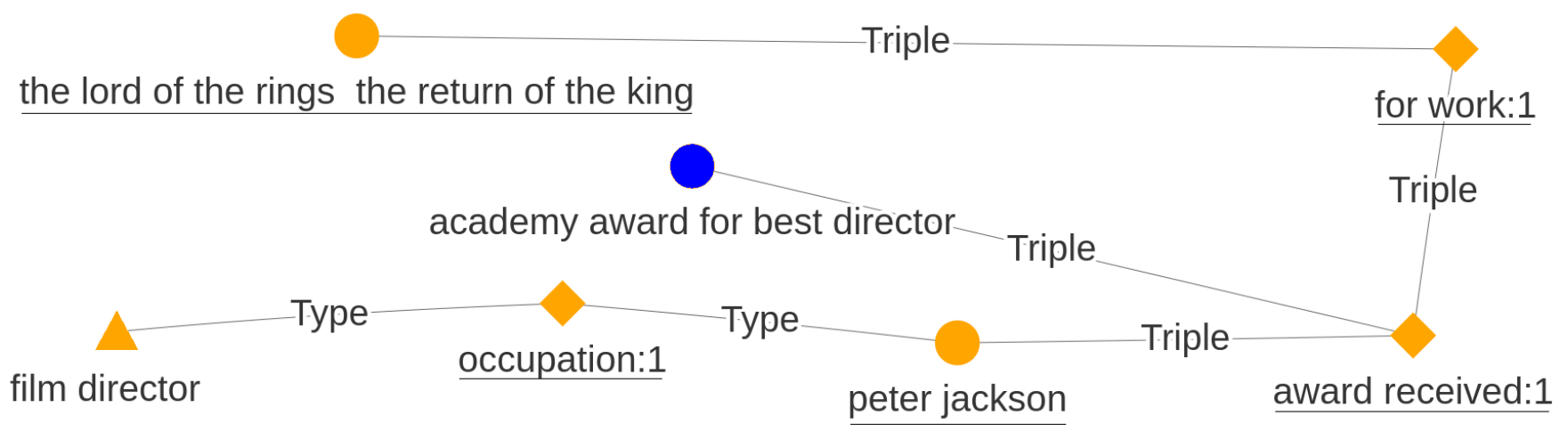

(b) Example GST for *What award did Peter Jackson receive for his work, "The Lord of the Rings: The Return of the King"?* in KG setup.

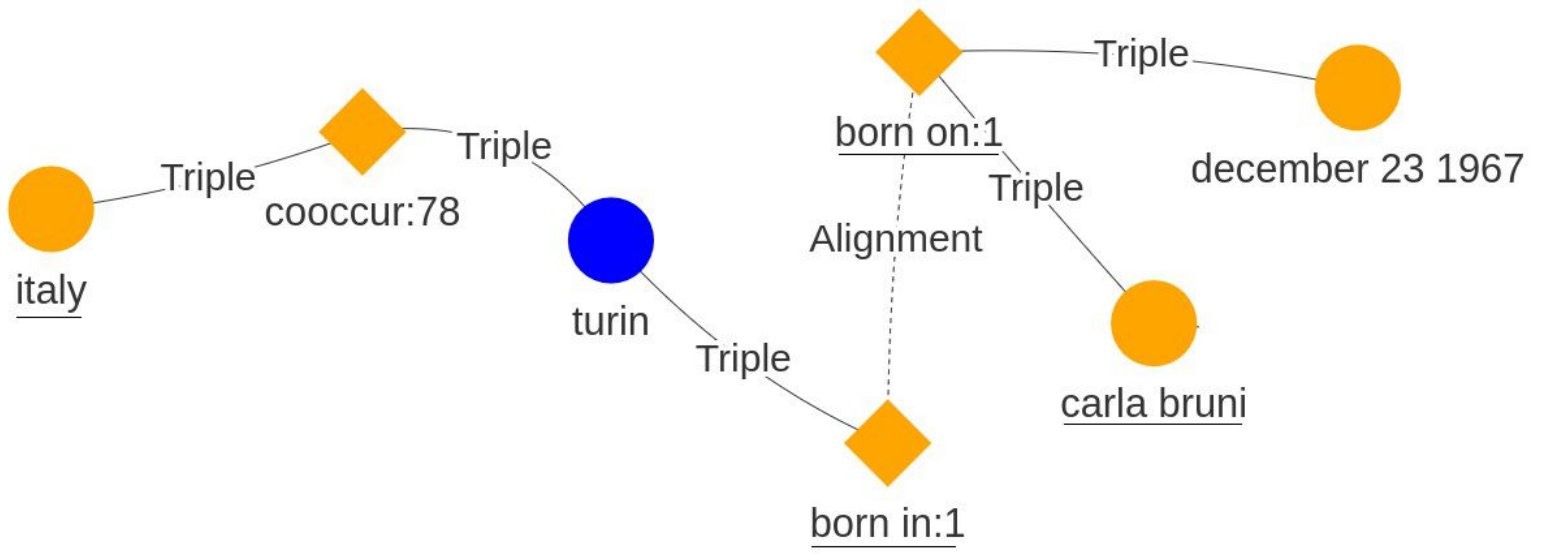

(c) Example GST for *Where in Italy was Carla Bruni born?* in Text setup.

The evaluation agent scored the LLM answer with 0 or 1 depending on whether the generated output gave the same entity as the gold answer (a proxy for Precision@1, our main metric in this paper). UNIQORN answers were also re-evaluated by GPT-4o in a similar manner for fairness. Note that this automated evaluation is geared for matching different alias names and surface phrases, whereas fuzzy string-matching or similar alternatives would often fail and human evaluation would be too time- and cost-intensive to cover for the numerous variations generated between various prompts, LLMs, and source-settings. The prompt was: *"Please compare the following answers and determine if the generated answer matches any of the gold answers semantically. For semantic matches (i.e., for getting a '1', the answers may not match exactly, but the two answers should refer to the same entity, date, or other constant. Respond with '1' if they match and '0' if they do not. Gold answers:* `[gold answers]`*; Generated answer:* `[generated answer]`*"*.

**Results**. Results on our dev set with 1000 questions, are in Table 17. All pairwise differences are statistically significant with p-values $\leq 0.05$, under the McNemar's test for paired binomial data, and significant improvements of UNIQORN over GPT-4o are marked with '*'. We found that GPT-4o produces far better performance than Phi-3 (0.213 vs. 0.036, for zero-shot; 0.249 vs. 0.061, for RAG over KG+Text). UNIQORN, in comparison, obtains the best results of 0.264 (KG+Text) and 0.273 (KG), outperforming even GPT-4o by a clear margin. Only for Text-QA, UNIQORN is inferior to GPT-4o. This is an exciting result, as the LLMs are pre-trained and fine-tuned on text corpora and other data that are much larger than the content that UNIQORN had access to. A zero-shot setting is inapplicable for UNIQORN as it has no parametric memory.

**Table 17**
Comparison of UNIQORN with LLMs on dev set (P@1).

| Method | Zero-shot | RAG (KG+Text) | RAG (KG) | Text |
|---|---|---|---|---|
| UNIQORN | N. A. | **0.264*** | **0.273*** | 0.043 |
| GPT-4o | **0.213** | 0.249 | 0.194 | **0.106** |
| Phi-3 | 0.036 | 0.061 | 0.098 | 0.012 |

Since GPT-4o was clearly the better LLM here, we conducted a more in-depth experiment with it, to assess the effect of parametric memory. In RAG mode, initial studies (manual investigation of sample outputs from row for GPT-4o in Table 17) revealed that, despite the restrictive instructions in the prompt, LLMs do occasionally give answers that are not present in the input prompt at all. To understand this phenomenon, we perturbed the retrieved top-$k$ evidence pieces, such that all surface forms of each of the ground-truth answer entities were replaced by the string *"Seqret Uniquorn"*. This way, the only faithful answer of the RAG-LLM should be a phrase that includes *"Seqret Uniquorn"* or at least *"Seqret"* or *"Uniquorn"*.

**Table 18**
Passage perturbation experiments with GPT-4o.

| Original evidence | RAG (KG+Text) | RAG (KG) | Text |
|---|---|---|---|
| Correct answers | 249 | 194 | 106 |
| LLM abstention message | 397 | 659 | 542 |
| Hallucinations | 32 | 27 | 28 |
| Incorrect answers from evidence | 293 | 120 | 296 |
| Context length exceeded | 29 | 0 | 28 |
| Total cases | 1000 | 1000 | 1000 |
| **Perturbed evidence** | **RAG (KG+Text)** | **RAG (KG)** | **Text** |
| "Seqret Uniquorn"+variants | 201 | 104 | 87 |
| Correct answers (match with gold) | 46 | 30 | 33 |
| LLM abstention message | 469 | 772 | 590 |
| Hallucinations | 30 | 21 | 23 |
| Incorrect answers from evidence | 223 | 73 | 238 |
| Context length exceeded | 31 | 0 | 29 |
| Total cases | 1000 | 1000 | 1000 |

Results of this evidence perturbation experiment are in Table 18. Answers containing variants of *"Seqret Uniquorn"* were returned 201, 104, and 87 times respectively for KG+Text, KG, and Text-RAG – compared to 249, 194, and 106 cases of correct answers in the original setting. An ideal result here would be close to 100% of the previously correct answer cases to be changed to a variant of *"Seqret Uniquorn"*. This indicates imperfect instruction-following capabilities for even powerful models like GPT-4o. Hand in hand, with perturbed passages, we would have expected zero LLM answers to match now with gold answers: but this still happens in 46, 30, and 33 cases for KG+Text, KG, and Text-RAG, respectively. This is clear evidence that parametric memory of the LLM kicks in here, despite clear instructions of not using it. More than 20 cases of **hallucinations** (an incorrect answer that was not contained in the evidence or the question) were observed in the outputs, which violates faithful QA. Notably, LLM abstention messages (`The given evidence does not contain the answer to the input question.`) were triggered much less often in the KG+Text scenario (397 times in original and 469 times in perturbed setting) compared to individual sources – showing that heterogeneous evidence indeed boosts answer coverage.

All in all, we think these are remarkable observations, as today's LLMs have billions to trillions of parameters and have gone through excessive pre-training and fine-tuning with huge corpora and human inputs. In contrast, UNIQORN is a transparent pipeline that faithfully computes answers solely from retrieved evidence. Although this experiment is small in scale and may not generalize for other knowledge-centric tasks or conversational settings, we view these results as demystifying the wide belief that LLMs with RAG solve the complex QA problem for good. All scripts (including prompts), intermediate results, and raw and processed data files for these LLM and RAG experiments with GPT-4o and Phi-3 are publicly available on UNIQORN's homepage and GitHub for transparency. To access these, please see the directory llm-expts/ inside the zip archive available at `https://qa.mpi-inf.mpg.de/uniqorn/UNIQORN_Code.zip`, or view the code on GitHub at `https://github.com/ajesujoba/UNIQORN/tree/main/llm-expts`.

## 7. Related work

### 7.1. QA over heterogeneous sources

Methods for heterogeneous QA can be broadly grouped as adopting one of the three following means [99]: (i) *early fusion*, where sources are merged early on via cross-source links in question-relevant context graphs [112, 111]; (ii) *late fusion*, where there are different pipelines for the individual sources, and they interact at later stages to fuse or rank candidate answers [11, 129, 131, 102, 45, 103]; or (iii) *unified representations*, where evidences from all sources are converted into a unified form [89, 24, 30, 86]. The last branch is recently emerging as a mechanism of choice in heterogeneous QA where evidences from all sources are verbalized as NL sequences [89]. This phenomenon can be attributed to the success of LLMs that can be harnessed for answer generation. UNIQORN also falls into the last bucket, but has a unique positioning of being the only work that investigates quite the opposite idea – inducing SPO structure on all heterogeneous evidences instead of verbalization. We show that when it comes to *reasoning* with *complex* intents, using graphs that leverage *explicit connections* between question-relevant evidences can be of critical value. Very recently, conversational QA [33, 24, 25] has started leveraging heterogeneous data like KGs, tables, and text, but the current models are tailored for incomplete questions with simple intents, and are not yet geared for tackling more complex cases: the same holds for domain-specific heterogeneous QA [108].

### 7.2. QA over knowledge graphs

The inception of large knowledge graphs like Freebase [15], YAGO [110], DBpedia [7], and Wikidata [124], gave rise to question answering over knowledge graphs (KG-QA) that typically provides answers as *single entities* or *entity lists* from the KG. KG-QA is now an increasingly active research avenue, where the traditional goal has been to translate an NL question into a structured query, usually in SPARQL syntax or an equivalent logical form, that is directly executable over the KG's RDF triple store containing entities, predicates, types and literals [128, 96, 122, 14, 22, 91, 65]. To circumvent the brittleness of SPARQL for complex intents, an alternative direction has used approximate graph search without explicit queries (UNIQORN adopts this philosophy) [122, 22, 25, 73, 139]. In some of these cases, the entire KG is cast into an embedded space for multihop reasoning [104, 105, 62], or the answer derivation workflow is cast into a sequence-to-sequence model [17, 116, 24]. The basic challenge in all cases is the same: bridging the vocabulary gap between phrases in questions and the terminology of the KG, i.e. mapping question tokens to KG items. Early work on KG-QA built on paraphrase-based mappings and question-query templates that typically had a single entity or a single predicate as slots [12, 120, 132]. This direction was advanced by [10, 9, 3, 59], including templates that were automatically learnt from graph patterns in the KG. Unfortunately, this dependence on templates prevents such approaches from coping with arbitrary syntactic formulations in a robust manner. This led to the development of deep learning methods with sequence models, and key-value memory networks [130, 131, 115, 62, 21, 66, 116, 17]. These have been successful on benchmarks like WebQuestions [12] and QALD [121]. However, these methods critically build on sufficient amounts of training data in the form of $\langle Q, A \rangle$ pairs. In contrast, the GST-based core of Uniqorn is unsupervised and needs neither templates nor training data.

*Complex question answering* is an area of intense focus in KG-QA now [61, 14, 96, 40, 122, 60, 67, 17, 116, 104], where the general approach is often guided by the existence and detection of substructures for the executable query. Uniqorn treats this as a potential drawback and adopts a joint disambiguation of question concepts using algorithms for Group Steiner Trees, instead of looking for nested question units that can be mapped to simpler queries. Approaches based on question decomposition (explicit or implicit) are brittle due to the huge variety of question formulation patterns (especially for complex questions), and are particularly vulnerable when questions are posed in telegraphic form (*oscar-winnning nolan films?*, has to be interpreted as *Which movies were directed by Christopher Nolan and won an Oscar award?*: this is highly non-trivial). Another branch of complex KG-QA rose from the task of knowledge graph reasoning (KGR) [29, 96, 35, 27, 140], where the key idea is given a KG entity (`Albert Einstein`) and a textual relation (*"nephew"*), the best KG path from the input entity to the target answer entity is sought. This can be generalized into a so-called multi-hop QA task [105, 104, 82] where the topic (question) entity is known and the question is assumed to be a paraphrase of the multi-hop KG relation (there is an assumption that *"nephew"* is not directly a KG predicate). Nevertheless, this is a restricted view of complex KG-QA, and only deals with such indirection or chain questions (*"nephew"* has to be matched with the `sibling` predicate followed by the `child` predicate in the KG), evaluated on truncated subsets of the full KG that typically lack the complexity of qualifier triples.

## 7.3. QA over text

Originally, in the late 1990s and early 2000s, question answering considered textual document collections as its underlying source. Classical approaches based on statistical scoring [98, 123] extracted answers as short text units from passages or sentences that matched most cue words from the question. Such models made intensive use of IR techniques for scoring of sentences or passages and aggregation of evidence for answer candidates. IBM Watson [45], a thoroughly engineered system that won the Jeopardy! quiz show, extended this paradigm by combining it with learned models for special question types. TREC ran a QA benchmarking series from 1999 to 2007, and revived it as the LiveQA [4] and Complex Answer Retrieval (CAR) tracks [38].

*Machine reading comprehension (MRC)* was originally motivated by the goal of whether algorithms actually *understood textual content*. This eventually became a QA variation where a question needs to be answered as a short *span of words* from a given text paragraph [97, 136], and is different from the typical fact-centric answer-finding task in IR. Exemplary approaches in MRC that extended the original single-passage setting to a multi-document one can be found in DrQA [19] and DocumentQA [26] (among many, many others). Traditional fact-centric QA over text, and multi-document MRC have recently emerged as a joint topic referred to as *open-domain question answering* [83, 31, 125, 78], also called the *retrieve-and-read* paradigm. Open-domain QA (ODQA) tries to combine an IR-based retrieval pipeline and NLP-style reading comprehension, to produce crisp answers extracted from passages retrieved on-the-fly from large corpora (see [20] for an overview). Uniqorn cannot outperform powerful SoTA models in this retriever-reader space like PathRetriever [6], and our focus was more on developing a seamless method that works over any source(s). Recall that ODQA models do not work over structured data, unlike Uniqorn.

## 7.4. Domain-specific QA

QA for vertical domains, like health, finance, sports, energy, and more, is a perfect use case for search over heterogeneous sources. For example, in biomedical QA, structured knowledge about proteins, diseases, drugs, etc. is crucial, and available in the form of specialized knowledge graphs or databases. While this provides core information on entities, the most interesting content is in free-text form in the literature, most notably, in PubMed articles, or in patients' clinical reports or even online health forums [53]. Thus, being able to tap into these different modalities is vital. The method of choice currently seems to be based on LLMs, but substantial customization is needed to encode sufficient domain expertise. Likewise, quality control for the training corpora is a challenging issue. For biomedial QA, [69] gives a survey on state-of-the-art methods. For finance QA, the company report [126] provides insights into the choice of heterogeneous data, the quality control for training, and the heavy engineering that are required for a viable solution. Investigating to what extent a graph-based approach like Uniqorn could potentially contribute to domain-specific QA (e.g., for domains beyond those of big enterprises, such as energy or climate) is a subject for future research.

## 8. Conclusions and Future Work

Answering complex factual questions over multiple, heterogeneous input sources requires a unified way of joining nuggets of evidence. Through our UNIQORN proposal, we show that computing Group Steiner Trees on noisy question-relevant context graphs, created on-the-fly by casting evidence from each source to a relaxed subject-predicate-object structure, is a viable solution. We demonstrate substantial performance gains of our model over the alternative paradigm of unification via verbalization, that loses vital relationships across pieces of evidence that are critical for faithful answering of complex intents. Further, use of Steiner Trees makes the answer derivation and provenance both transparent to an end-user, an open concern for sophisticated models based directly on LLMs. In a small-scale but insightful experiment, we also show that complex and heterogeneous factual QA is still not a solved problem for the latest LLMs, as UNIQORN outperforms these models both on zero-shot results as well as in a comparable RAG setup. UNIQORN works over heterogeneous sources, but always provides crisp entities or phrases as answers. A key future direction would be to involve a different kind of heterogeneity: allowing for answers at different levels of granularity.

## Acknowledgements

The authors would like to thank Philipp Christmann from the Max Planck Institute for Informatics (MPII) for helping in experiments involving the CLOCQ API. We also acknowledge Magdalena Kaiser from the MPII and Zhen Jia from the South West Jiao-Tong University of China, for useful inputs at various stages of this work.
**Funding:** This work was supported by the ERC Synergy Grant 610150 (imPACT).

## CRediT authorship contribution statement

**Soumajit Pramanik:** Investigation, Data curation, Methodology, Resources, Software, Validation, Visualization, Writing – review & editing. **Jesujoba Alabi:** Data curation, Methodology, Resources, Software. **Rishiraj Saha Roy:** Formal analysis, Investigation, Methodology, Project administration, Supervision, Validation, Visualization, Writing – original draft, Writing – review & editing. **Gerhard Weikum:** Conceptualization, Formal analysis, Investigation, Methodology, Project administration, Supervision, Validation, Writing – review & editing.